# Phase amplification microscopy with femtometer-level accuracy

Nansen Zhou[1,†], Qihang Zhang[1,†], Ting Huang[1], Helios Y. Li[2], Chenjia Xi[1], Jiawen You[1,3], Jinsong Zhang[4], Yujie Nie[1], Chaoran Huang[5], Zhaoli Gao[1], Jinlong Zhu[4], Qiwen Zhan[1,6], Jianbin Xu[5], Nicholas X. Fang[2,7,8,*], and Renjie Zhou[1,5,*]

[1]Department of Biomedical Engineering, The Chinese University of Hong Kong, Shatin, New Territories, Hong Kong SAR, China

[2]Department of Mechanical Engineering, The University of Hong Kong, Hong Kong SAR, China

[3]Department of Chemical and Biological Engineering, The Hong Kong University of Science and Technology, Clear Water Bay, Kowloon, Hong Kong SAR, China

[4]State Key Laboratory of Digital Manufacturing Equipment and Technology, Huazhong University of Science and Technology, Wuhan 430074, China

[5]Department of Electronic Engineering, The Chinese University of Hong Kong, Shatin, New Territories, Hong Kong SAR, China

[6]Zhejiang Key Laboratory of 3D Micro/Nano Fabrication and Characterization, Department of Electronic and Information Engineering, School of Engineering, Westlake University, Hangzhou, Zhejiang, China

[7]HK Institute of Quantum Science & Technology, The University of Hong Kong, Pokfulam Road, Hong Kong SAR, China

[8]Materials Innovation Institute for Life Sciences and Energy (MILES), HKU-SIRI, Shenzhen, China

*Corresponding authors: nicxfang@hku.hk; rjzhou@cuhk.edu.hk

[†]These authors contributed equally to this work.

**ABSTRACT**

We demonstrate a major breakthrough in laser interferometry and microscopy achieving femtometer-level measurement accuracy and beyond, termed Phase Amplification microscopy (Φ-Amp). By exploiting the native silicide substrate as a phase cavity, our phase-gain theory predicts that weak sub-atomic phase signals can be magnified over 1000-fold, thus bypassing the shot-noise limit. We experimentally achieved a 158.2-fold phase gain for graphene in ambient air, corresponding ~ 730 femtometer accuracy. To fully unleash the potential of Φ-Amp for atomic fabrication and quantum measurement, we quantified interlayer spacing differences between AB-stacked and 30°-twisted bilayer graphene to be ~ 0.77 Å and further detected atomic impurities and defects on large atomic structures. As the first wide-field metrology tool, we envision Φ-Amp may accelerate the scaling up of atomic quantum devices.

**MAIN**

Recent advances in stacked two-dimensional (2D) quantum materials have driven extensive research for engineering electronic properties at the subatomic scale[1-5]. In stacked 2D materials, such as twisted bilayer graphene (tBLG)[2, 6], the interlayer rotation angle serves as a powerful knob to induce phenomena ranging from unconventional superconductivity to quantum anomalous Hall effects, thus promising a new generation of high-mobility, low power atomic devices[7-10]. Yet, the practical realization of these devices hinges critically on the precise characterization of their structural integrity: interlayer spacing, defects, and contaminations, all profoundly influence the device performance[11]. As exemplified by the 2023 International Roadmap for Devices and Systems (IRDS)[12], a pressing metrology gap exists, namely, the ability to correlate atomic-scale structural information with physical properties at a throughput compatible with beyond-CMOS manufacturing (**Supplementary Fig. S1**)[12, 13]. The existing metrology toolbox, however, is caught in a fundamental trade-off. Scanning probe and electron microscopies[13-17] offer sub-Ångström feature resolution, but are inherently serial and slow, rendering them impractical for wafer-scale inspection. On the other hand, far-field optical techniques, such as bright-field microscopy, scatterometry, and interferometric scattering microscopy[12, 13, 18, 19], are rapid and non-destructive, yet their sensitivity falls short of the atomic scale.

Laser interferometry offers a tantalizing pathway for detection of weak signals, as profoundly demonstrated by gravitational wave observation by the Laser Interferometer Gravitational-wave Observatory (LIGO)[20]. In microscopy, the seminal work of Zernike on phase contrast microscopy[21-23] and the subsequent development of quantitative phase microscopy (QPM) have enabled label-free profiling of thin structures with nanometric axial precision[24-33]. In QPM, the measurement accuracy is governed by the phase signal-to-noise ratio (SNR) and the fidelity of the reconstruction model. Considerable efforts have therefore focused on noise suppression, including common-path interferometry detection[34, 35], illumination coherence reduction[36-38], dynamic range expansion[39], and high-well-capacity detection[40, 41]. Yet, despite these advances, the lowest reported spatial phase noise under ambient conditions remains at $\sim 10^{-4}$ rad[35], even after frame summing that severely compromises imaging speed, and this translates to Ångstrom-level accuracy for measuring single-layer atomic structures. Further scaling down the phase noise is extremely challenging, due to air disturbances, mechanical vibrations and ultimately, due to the fundamental limit of photon shot noise.

Here, we introduce Phase Amplification microscopy (Φ-Amp), a technique that exploits the native silicon-compatible substrate as a phase cavity to physically amplify the weak phase signal of a 2D material before detection, without amplifying noise. Our phase-gain theory predicts that a sub-$10^{-5}$ rad phase shift can be amplified to a detectable level of $10^{-3}$ rad, outperforming the noise floor of conventional QPM systems. This is achieved by exploiting the native substrate compatible with the silicon platform as a phase cavity and choosing appropriate operating wavelengths (**Fig. 1a**). While the reflected intensity approaches a minimum at resonance, reminiscent of the nulling principle in Bracewell interferometry[42], our off-axis interferometric detection faithfully recovers the amplified phase with exceptional fidelity. We experimentally achieved 158.2-fold phase amplification on graphene, leading to femtometer-level thickness measurement accuracy. This unprecedented sensitivity and accuracy allows us to quantify, for the first time, the interlayer spacing difference between AB-stacked and 30°-twisted bilayer graphene to be ~ 0.77 Å, in excellent agreement with density functional theory (DFT) simulations. Furthermore, the wide-field capability of Φ-Amp enables rapid mapping of atomic-scale defects and impurities across large areas, offering a practical route for high-throughput quality screening in atomic device fabrication.

Beyond graphene, the phase-gain concept is generalizable to other 2D materials and substrate architectures. With the availability of attosecond sources[44] and the ongoing advances in cavity engineering inspired by $10^{-19}$ meter level accuracy of LIGO[43] , we envision that Φ-Amp can be further scaled to sub-attometer accuracy, potentially unveiling the subtle electronic dynamics that govern quantum materials.

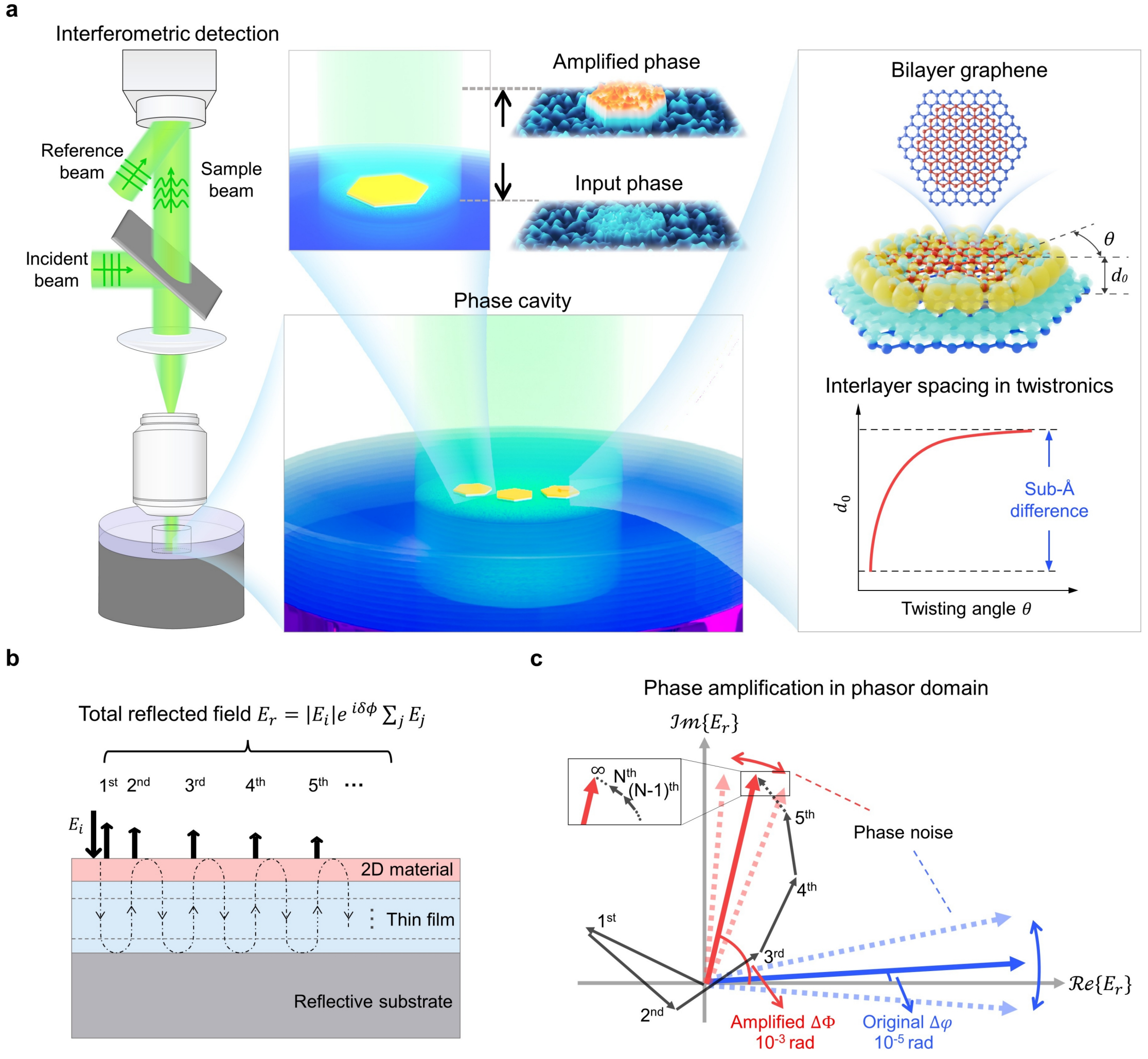

**Fig. 1. Working principle of Φ-Amp in amplifying weak phase signals and applications in wide-field mapping of sub-atomic structures**. **a,** Schematic design of Φ-Amp based on a reflection-mode laser interferometric microscope. A zoomed-in image over the sample region shows the phase cavity that amplifies the weak phase signals. Using graphene as an example, the graphene structure is overwhelmed by the phase noise in the original input phase map (bottom), while the graphene is revealed with high contrast (top) with the phase amplification by the cavity. Furthermore, Φ-Amp enables the quantification of interlayer spacing differences of twisted bilayer graphene. **b,** Physical mechanism of phase amplification. The total reflected field ($E_r$) from a multilayer structure is the superposition of an infinite number of partial waves, which can be represented in the phasor domain as shown in **c**. In a phase cavity, the phase of $E_r$ can reach the maximum value $\Delta\Phi$ (red). For the interlayer spacing difference with $10^{-5}$ rad level original phase $\Delta\varphi$ (blue), we can amplify it to a detectable phase at $10^{-3}$ rad level (red). During the amplification process, the phase noise $\delta\phi$ stemming from the incident wave (denoted by dotted red and blue arrows) is not amplified but can slightly influence the amplified phase value.

**Phase-gain theory, modeling, and experimental verification**

**Figure 1** demonstrates the working principle of phase amplification microscopy, which is based on a reflection-mode laser interferometric microscope as shown in **Fig. 1a**. A collimated laser beam passes through the tube lens and the objective lens before illuminating the thin sample. Upon reflection, the sample scattered field is collected by the same objective lens and directed to the camera, where it interferes with a reference beam to form an interferogram, from which the phase map is retrieved. Due to the layered structure of the sample, multiple refractions and reflections occur on the interfaces as illustrated in **Fig.1b**. By tuning the wavelength or substrate thicknesses to satisfy the phase resonance condition, the field summation results in an enhanced phase as shown in **Fig. 1c**. To quantify the reflected field, **Fig. 2a** presents a multilayer structure composed of a sample layer (1st layer, i.e., the atomic structure), a reflective substrate ($N^{th}$ layer), and intermediate layers denoted as thin film. Each layer is characterized by its refractive index $\widetilde{n}_i$ and thickness $H_i$. Assuming a monochromatic incident plane wave in free space with a wavelength of $\lambda$, the total reflected complex field can be obtained by modeling light propagation in the multilayer structure under the effective medium approximation, as widely used in modeling metamaterials, atomic structures, etc[45-47]. In the model, we first construct an equivalent reflection interface between the sample and all the underlying layers (2nd to $N$th layer). At this interface, in analogy to the Fresnel reflection coefficients, we define an equivalent reflection coefficient $r_1^N$, which is related to the effective refractive indices $\widetilde{n}_{1,eff}^{(N-1)}$ and $\widetilde{n}_{N,eff}^{(N-1)}$. Through a recursive relationship between all the equivalent reflective interfaces, we connect $r_1^N$ with the corresponding equivalent reflection coefficients $r_{m-1,N}$ ($m$ = 2, …, $N$-1). Using $r_{1,N}$, we obtain the phase delay of the total reflected field ($\Delta\Phi$), relating to the refractive index $\widetilde{n}_i$ and thickness $H_i$ of each layer and $\lambda$. Similarly, the phase of the sample on a bare wafer is defined as the original phase ($\Delta\varphi$). Finally, we define the phase gain $G$ to evaluate the phase amplification capacity for weak signal samples under different phase cavity structure,

$$G\{\widetilde{n}_i; H_i; \lambda\} = \left|\frac{\partial(\Delta\Phi)}{\partial(H_1)} \Big/ \frac{\partial(\Delta\varphi)}{\partial(H_1)}\right| \tag{1}$$

where $i = 1,2, \ldots, N$. Note that when measuring atomic structures with sample thickness $H_1 \ll 0.01\lambda$, both $\Delta\Phi$ and $\Delta\varphi$ exhibit linear relationships with $H_1$ after the Taylor expansions, thus $G$ is not related to $H_1$. As shown in Eq. (1), $G$ varies with wavelength $\lambda$ and the thicknesses of the individual thin film layers. At certain wavelengths or substrate conditions, $G$ reaches maximum values, which are referred to as the phase resonance conditions. If the phase noise, such as laser speckle, is presented in the incident field before entering the cavity, it will not be amplified by the phase cavity (refer to discussions and detailed derivations in **Supplementary Note 1**).

To experimentally verify the phase amplification, we built a reflection-mode laser interferometric microscopy system with a 20× objective lens (EC Epiplan-Neofluar 20×/0.50 Pol M27, Carl Zeiss AG). The system design is illustrated in **Fig. 2b**, and the physical picture of the imaging system is shown in **Supplementary Methods**. After mounting the sample (zoomed-in image in **Fig. 2b**), we captured interferograms and retrieved the phase maps. To extract quantitative phase values, a mask-based standardized data analysis workflow was developed, as illustrated in **Fig. 2c**. The phase maps are first de-tilted to remove global phase gradients, followed by edge detection to identify the region of interest. The enclosed regions are then filled to generate binary masks, which are applied to segment the MLG regions from the background in each phase image. The resulting masked phase maps are subsequently used to extract quantitative data and perform statistical analysis.

We applied the commonly presented $SiO_2$ layer on silicon wafers to realize a single-layer phase cavity for amplifying the phase signal of monolayer graphene (MLG). We simulated *G* as a function of $\lambda$ and $SiO_2$ thickness ($H_2$) as shown in **Fig. 2d**, which reveals that *G* oscillates as a function of $H_2$ with a period of $0.5\lambda/n_2$, and displays an asymmetry caused by sample absorption and loss-enhanced phase amplification, as detailed in **Supplementary Methods** and **Note 2**. Therefore, the maximum *G* can be reached by either selecting an appropriate wavelength ($\lambda$) or designing the thickness of $SiO_2$. At the widely used laser wavelength of $\lambda$ = 532 nm, when the phase resonance condition is satisfied (e.g., $H_2$ = 285.5 nm in **Fig. 2d**), *G* reaches a maximum value of 31.2 on MLG that results in a maximum phase value of 87.8 mrad (**Fig. 2f**). To experimentally validate the dependence of the phase on $H_2$, we prepared MLG samples and transferred them to Si wafers with 5 different $SiO_2$ layer thicknesses (sample preparation and measurement conditions are provided in **Supplementary Methods**). After extracting the phase data and performing statistical analysis, we found the phase values in the MLG regions agree with predicted values from our phase-gain model (**Fig. 2f**). The standard deviations of the phase values for the 500 phase maps (blue error bars) are small, which indicates the system has a high stability, and the values agree with predicted values from the photo shot noise model (refer to section on **Accuracy limit and verification**). For the sample with a $SiO_2$ layer thickness of 285.8 nm (measured using ellipsometry), the amplified phase is $81.2 \pm 0.1$ mrad and $G$ = 28.9, thus achieving around 30 times amplification as expected. Consequently, the phase signal-to-noise ratio (SNR) is enhanced by the same factor, thus leading to a significant enhancement of the image contrast as quantified by the contrast-to-noise ratio (CNR) to $7.11 \pm 0.06$ dB. CNR is defined as $10\log10(\Delta\Phi/\delta\Phi_{bg})$, where $\delta\Phi_{bg}$ is the background noise. The results further confirm that the phase noise is not amplified. In addition, our phase-gain model can reproduce previous experimental results based on enhancing the intensity contrast of MLG[48] (**Supplementary Note 2**).

To further validate the wavelength dependence of the phase response, we performed wavelength-scanning measurements on an MLG sample on a fixed $SiO_2$ thickness of $322 \pm 5$ nm using a USB camera (Alvium 1800 U-2460m**,** Allied Vision Technologies GmbH). The wavelength scanning was achieved using a tunable wavelength laser system consisting of a supercontinuum white-light laser (SuperK EXTREME EXR-15, NKT Photonics A/S) coupled with a tunable wavelength filter based on Acousto-Optic Tunable Filters (AOTF) (SuperK SELECT VIS-nIR, NKT Photonics A/S). By tuning the illumination wavelength from 515 nm to 695 nm in steps of 5 nm, we found that the measured phase in the MLG region reaches its maximum value of $69.3 \pm 1.0$ mrad at $\lambda$ = 605 nm (**Fig. 2g**) with a CNR of $8.40 \pm 0.59$ dB. Away from the resonance, the phase response decreases and approaches zero at $\lambda$ = 670 nm and becomes negative for $\lambda > 670$ nm. Through fitting the experimental results, we found that the thickness of $SiO_2$ in the measured region is around 327 nm. Moreover, the small standard deviations across repeated measurements at each wavelength indicate that wavelength tuning does not introduce additional phase noise. Additionally, we conducted wavelength-scanning measurements on the same sample using a 50× objective lens (TU Plan Fluor Epi P 50×/0.80, Nikon Corp.). The measured phase-wavelength dependence relationship was consistent with that obtained with the 20× objective lens. For objects whose dimensions are larger than the system's diffraction limit, these results indicate that the phase gain is independent of the objective magnification and numerical aperture. It also demonstrates the reproducibility and system-level robustness of the wavelength-scanning phase measurement across different imaging configurations (see **Supplementary Note 3**).

The phase cavity can be generalized to other commonly used wafers, such as $Si_3N_4$/Si and SiC/Si (refer to more details in **Supplementary Note 4**). With a multi-layer phase cavity, similar to high-quality factor cavities[49], the phase gain could reach higher. Guided by the phase-gain theory, our simulations in **Fig. 2e** demonstrate that a phase gain of over $10^3$ can be achieved using a simple double-layer phase cavity based on the $SiO_2/Si_3N_4$/Si structure. As a proof-of-concept demonstration in **Fig. 2h**, we devised a double-layer phase cavity with the measured $SiO_2$ thickness of 164.6 nm and $Si_3N_4$ thickness of 69.3 nm, which indicates a monotonic phase enhancement towards short wavelength. By tuning the illumination wavelength from 528 nm to 535 nm in steps of 0.5 nm, we found that phase wrapping effect induced by impurities begins to appear at $\lambda$ = 530.0 nm, resulting a decay of mean value in the MLG region when the wavelength goes smaller than 529.0 nm (refer to further discussions in **Fig. 5**). Therefore, we report a conservative phase gain value $G$ = 158.2 at $\lambda$ = 530.0 nm to avoid the complicated phase unwrapping process. More results about the wavelength dependence of MLG on the double-layer cavity are included in **Supplementary Note 5**. Due to the limited power of the supercontinuum laser, we acquired 100 images with an exposure time of 2.5 seconds each to calculate the average and standard deviation of the MLG phase values at each wavelength. It should be noted that the absorption in the cavity was so high that we had to use a 50× objective lens to collect more light to maintain the fringe contrast. To realize over $10^3$ phase amplification, it requires more precise and high-quality film fabrication and a narrow-linewidth laser with a bandwidth of less than 0.01 nm. A higher laser power or a more sensitive camera is also practically required to compensate for the absorption in the cavity (**Supplementary Notes 5 & 7**). Details about the phase cavity thickness characterization, phase gain simulation, and corresponding experimental results are provided in **Supplementary Notes 5 & 6**.

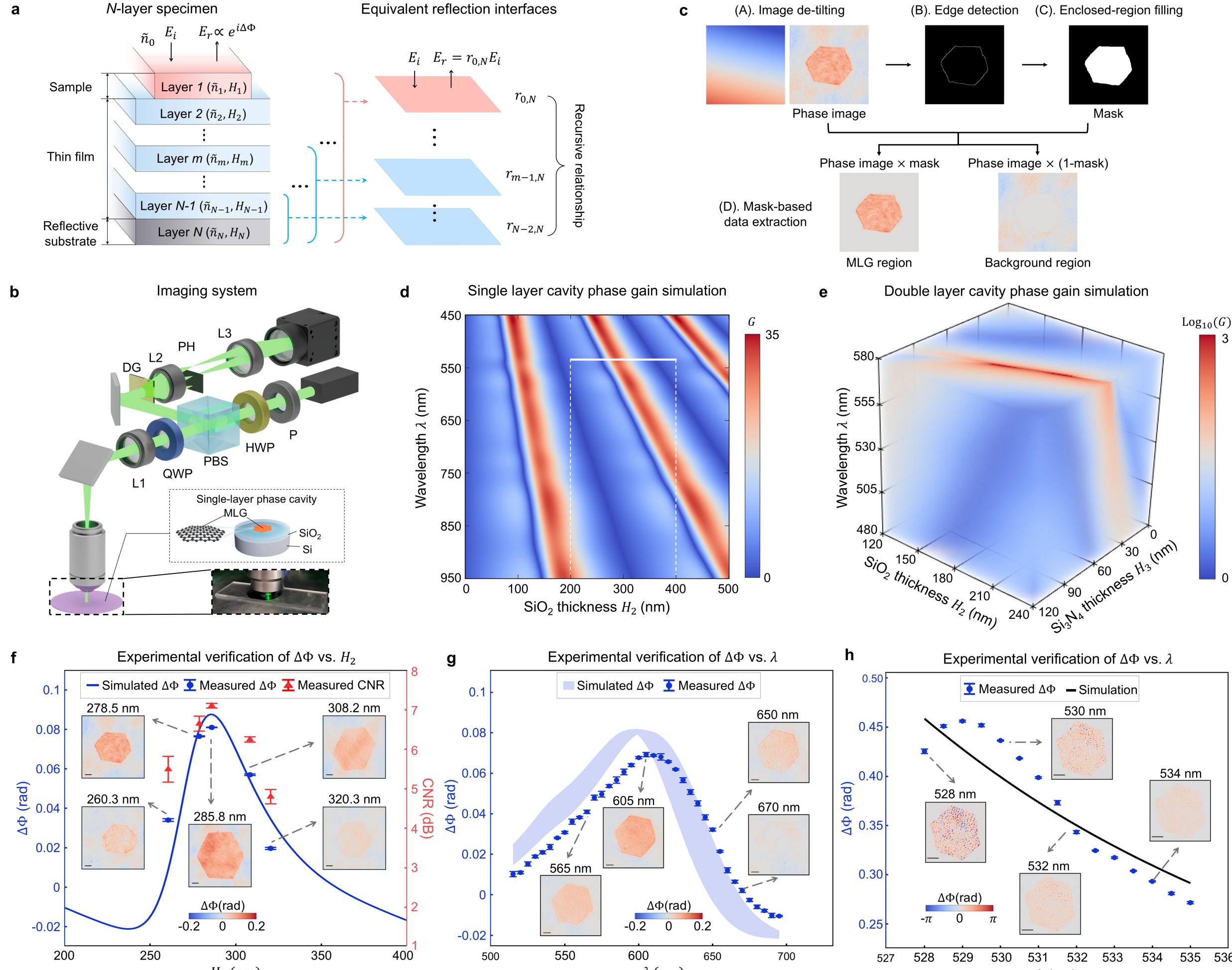

**Fig. 2. Phase gain simulation and experimental verification of phase amplification on monolayer graphene**. **a,** Modeling of the total reflected field from an $N$-layer specimen by constructing equivalent reflection interfaces. **b,** The imaging system of Φ-Amp. P: polarizer; HWP: half-wave plate; QWP: quarter-wave plate; PBS: polarization beam splitter; L1-L3: lens; DG: diffraction grating; PH: pinhole. The bottom right corner of **b** shows the design and physical pictures of the single-layer phase cavity. **c,** Illustration of the standardized data analysis workflow. (A). Image de-tilting of the phase map. (B). Edge detection. (C). Enclosed-region filling for mask generation. (D). Resulting phase maps of the MLG region and background region for subsequent data extraction and analysis. **d,** Phase gain simulation as a function of wavelength λ and $SiO_2$ thickness $H_2$. At a fixed wavelength of 532 nm, the relation between $G$ and $H_2$ is periodic, and the maximum $G$ is 31.2 when $H_2 = 285.5$ nm. **e,** Phase gain simulation as a function of $SiO_2$ thickness $H_2$, $Si_3N_4$ thickness $H_3$ and wavelength λ. At a fixed wavelength of 532 nm, the maximum $G$ is 1306 when $H_2 = 178.7$ nm and $H_3 = 66.6$ nm. We used the calculated refractive index of graphene (**Supplementary Note 9**) in the simulation plot **d** and **e**. **f,** ΔΦ vs. $H_2$ over the white line region in **d**. When $H_2 = 285.8$ nm, the measured phase has a maximum value of 81.2±0.1 mrad. **g,** ΔΦ vs. λ for the single layer cavity. The band represents the simulation results using $SiO_2$ thickness ranging from 317 nm to 327 nm. The wavelength scanning range is from 515 nm to 695 nm with an interval of 5 nm. **h,** ΔΦ vs. λ for the double-layer cavity. The black curve represents the simulation results using the measured $H_2$ = 164.6 nm and $H_3$ = 69.3 nm. The phase wrap gradually appears for the wavelength shorter than 530.0 nm, resulting a turning point of mean phase at λ = 529.0 nm. Insets in **f**, **g** and **h** show some typical phase maps of MLG samples with a scale bar: 5 μm.

**Accuracy limit and verification**

We quantify the measurement accuracy of Φ- Amp through a rigorous error-propagation analysis of our phase- gain model, yielding

$$\sigma_H = \alpha \frac{\sigma_\phi \lambda}{G}, \tag{2}$$

where $\sigma_\phi$ is spatial phase noise, and $\alpha = \left(4\pi|\mathrm{Re}\{\frac{\tilde{n}_1^2 - n_0^2}{\tilde{n}_{sub}^2 - n_0^2}\}|\right)^{-1}$ is a constant related to the sample refractive indices of sample $\tilde{n}_1$, surrounding medium $\tilde{n}_0$ and substrate $\tilde{n}_{sub}$. Detailed derivations of accuracy $\sigma_H$ are provided in **Supplementary Note 7.** In the absence of environmental disturbances, the ultimate limit of phase fluctuation $\sigma_H$ is set by photon shot noise: $\sigma_\phi = 1/\sqrt{mN_{eff}}$, where $m$ represents the effective number of pixels within a diffraction spot and $N_{eff}$ denotes the effective electron-well-capacity[41].

For our system employing a camera with a high full well capacity of $2\times10^6$ electrons (Q-2HFW-Hm/CXP-6, Adimec), we estimate $\sigma_\phi = 2.6\times10^{-4}$ rad. For the case of double-layer $SiO_2/Si_3N_4/Si$ phase cavity and monolayer graphene ($G$ = 1306 and $\alpha$ = 0.307, refractive index determined by density functional theory and many- body perturbation theory[50]), this translates into an accuracy of 32 fm, a remarkable figure that underscores the potential of the phase- cavity approach.

In practice, however, the achievable accuracy is governed by the actual spatial phase noise of the system. Using the lowest spatial phase noise of around $2.5\times10^{-4}$ rad in transmission-mode laser interferometric microscopy [34, 35, 41], we estimate the accuracy to be ~100 pm for profiling MLG. With Φ- Amp, this accuracy can be dramatically enhanced: rather than struggling to suppress noise by an order of magnitude as in LIGO, which may require kilometer- scale underground facilities or space- based isolation, we simply amplify the signal by $10^2$ or more. The details of accuracy estimation in transmission-mode systems are provided in **Supplementary Note 7.**

To test the measurement accuracy limit in Φ-Amp as illustrated in **Fig. 3a**, we first evaluated the system's phase noise by acquiring a stack of 500 interferograms of a sample-free region within 1 second. The representative spatial phase noise maps under single-frame measurement and temporal averaging are shown in the upper left and the lower right of **Fig. 3b(i)**, respectively. We quantified the measurement accuracies under four cases in **Fig. 3b(ii)**: for the above experimentally realized phase cavities with $G = 28.9$ and 158.2, the accuracies are 0.1 Å, and 1.9 pm, respectively; for the case of using temporal averaging at $G = 158.2$, the accuracy is ~ 730 fm; while the accuracy is only 3 Å for regular phase imaging (without phase gain) that is consistent with reported values[41]. **Figure 3b(iii)** further illustrates the improvement in accuracy by increasing the number of frames ($P$). Using the $SiO_2/Si_3N_4/Si$ double-layer phase cavity with $G = 1306$, a fm-level accuracy can be achieved by using a special narrow linewidth laser with a bandwidth of < 0.01 nm and a temperature fluctuation of < 0.2 °C, we predict that **femtometer-level accuracy** is readily achievable—a prospect we detail in **Supplementary Note 7**.

Note that the measurement accuracy mentioned here is sometimes referred to as axial/vertical resolution. Detailed descriptions of the differences in repeatability, sensitivity, precision, accuracy, and spatial resolution are illustrated in **Supplementary Methods**. The analysis of noise composition is illustrated in **Supplementary Note 8.**

The bright-field images and phase images under a single-layer phase cavity and a double-layer phase cavity are shown in **Fig. 3c** and **3d**, respectively. Using the calculated refractive index and applying our reconstruction model to the phase maps of MLG samples with a single-layer cavity of $G = 28.9$ (**Fig. 3c(iii)**) and a double-layer cavity of $G = 158.2$ (**Fig. 3d(iii)**), we obtained the thickness maps. Masks labeling the MLG regions are applied to thickness maps and their height histograms are plotted in **Fig. 3c(iv)** and **Fig. 3d(iv)**, respectively. For the single-layer phase cavity, we quantified MLG thickness to be 4.18 Å, and for the case of double-layer phase cavity, we quantified MLG thickness to be 4.288 Å. Notably, the phase image acquired with the double-layer cavity in **Fig. 3d(iii)** exhibits more pronounced spatial fluctuations, which manifest as a broader thickness distribution in the histogram. This increased sensitivity, far from being a drawback, reveals the presence of sub-Ångström surface variations—likely arising from impurities or substrate inhomogeneities—that are rendered visible by the high gain of the double-layer cavity. Indeed, these features are further investigated in **Fig. 5c,** where we correlate them with atomic-scale defects and contaminants.

In **Supplementary Note 5**, we show a cleaner another MLG sample on double-layer cavity with a G=101.7, yielding a more uniform thickness distribution, which confirms that the fluctuations observed in Fig. 3d are sample-specific rather than instrumental artifacts.

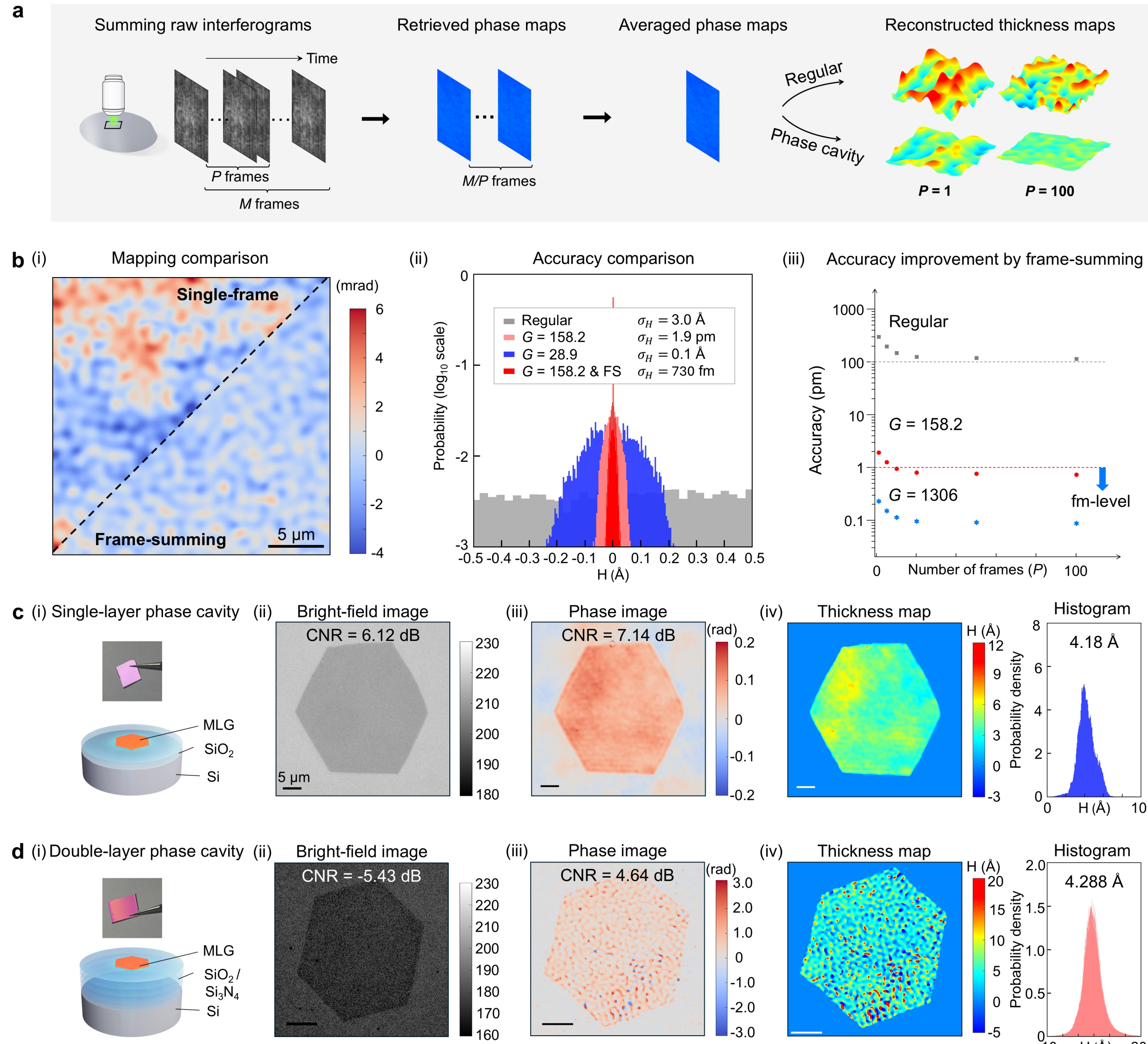

**Fig. 3. Measurement accuracy verification**. **a**, Workflow for evaluating the measurement accuracy. **b**, (i) Comparison of phase noise under single-frame measurement and applying frame summing; (ii) Comparison of measurement accuracy: regular phase imaging, with cavities ($G$ = 28.9 and $G$ = 158.2), and with cavity ($G$ = 158.2) and frame summing (FS). (iii) Accuracy improvement through frame summing. **c** and **d**, Imaging results for MLG samples under a single-layer phase cavity **c** and a double-layer phase cavity **d.** (i) The design and physical pictures of the single-layer phase cavity; (ii) The bright field image of MLG, (iii) The phase image of MLG, and (iv) the thickness map and the corresponding histogram of MLG sample.

## Profiling patterned atomic structures

The diffraction-limited lateral resolution of our imaging system is 665 nm when NA = 0.8. The fabrication process of 2D materials can leave behind residues and contaminants from polymers or etching solutions. To demonstrate the capability of Φ-Amp for high-throughput smoothness characterization of 2D material samples, we used a 20× objective lens and applied image stitching. **Figure. 4a** shows the reconstructed thickness map of MLG flakes over the whole FOV of 400 μm × 400 μm. We compared the region enclosed by the black box in **Fig. 4a** with the AFM mapping result, which proved that we could detect the residual particles left from sample preparation. We further selected three MLG regions **(i) - (iii)** in **Fig. 4a** and extracted their thickness maps and corresponding histograms, as shown in **Fig. 4b**. Analysis of these thickness maps yields an average MLG thickness of 4.70 Å, in good agreement with our previous measurements.

To test the profiling capability of Φ-Amp on atomic structures, we designed a variety of patterns on an MLG film and fabricated them using electron beam lithography (EBL), as illustrated in **Fig. 4c**. Detailed sample preparation procedures are described in **Supplementary Methods**. **Figure 4d(i)** and **(ii)** show the retrieved thickness maps of the same CUHK logo (linewidth of 1.4 μm) measured with Φ-Amp and AFM in the tapping mode (MultiMode 8-HR Bruker Corp.) and their histograms, respectively. From the histogram of Φ-Amp, we quantified the thickness of the CUHK pattern as 4.84 Å, while AFM shows 8.2 Å. The measurement results are distinct because the two techniques rely on fundamentally different working principles. AFM is based on the detection of tip–sample interactions[51], while Φ-Amp is based on the measurements of optical phase delay. By comparison, both Φ-Amp and AFM show consistent sample unevenness likely due to imperfections in fabrication[52,53], while Φ-Amp shows a much higher CNR of 5.9 dB vs. 2.4 dB by AFM, thanks to the amplification of the phase signal. Notably, Φ-Amp is close to $10^6$ times faster than AFM (2 ms vs. 17 minutes at the same pixel resolution of 60 nm). We summarized the attributes of Φ-Amp in **Supplementary Fig. S2**.

We further evaluated the spatial resolution of Φ-Amp by imaging resolution line pairs with periods of 720 nm and 1.5 μm, respectively. The reconstructed thickness maps obtained from Φ-Amp and AFM are shown in **Fig. 4e(i)** and **(ii)**, respectively. By plotting profiles along the white lines in **Fig. 4e(i)**, the spacings of both line pairs were quantified to be 720 nm. The results show that the line pairs can be well resolved, matching our system's specifications. From the 375 nm-wide features in **Fig. 4e(i)**, the due to the diffraction limit that attenuates the phase gain to around 16.5.

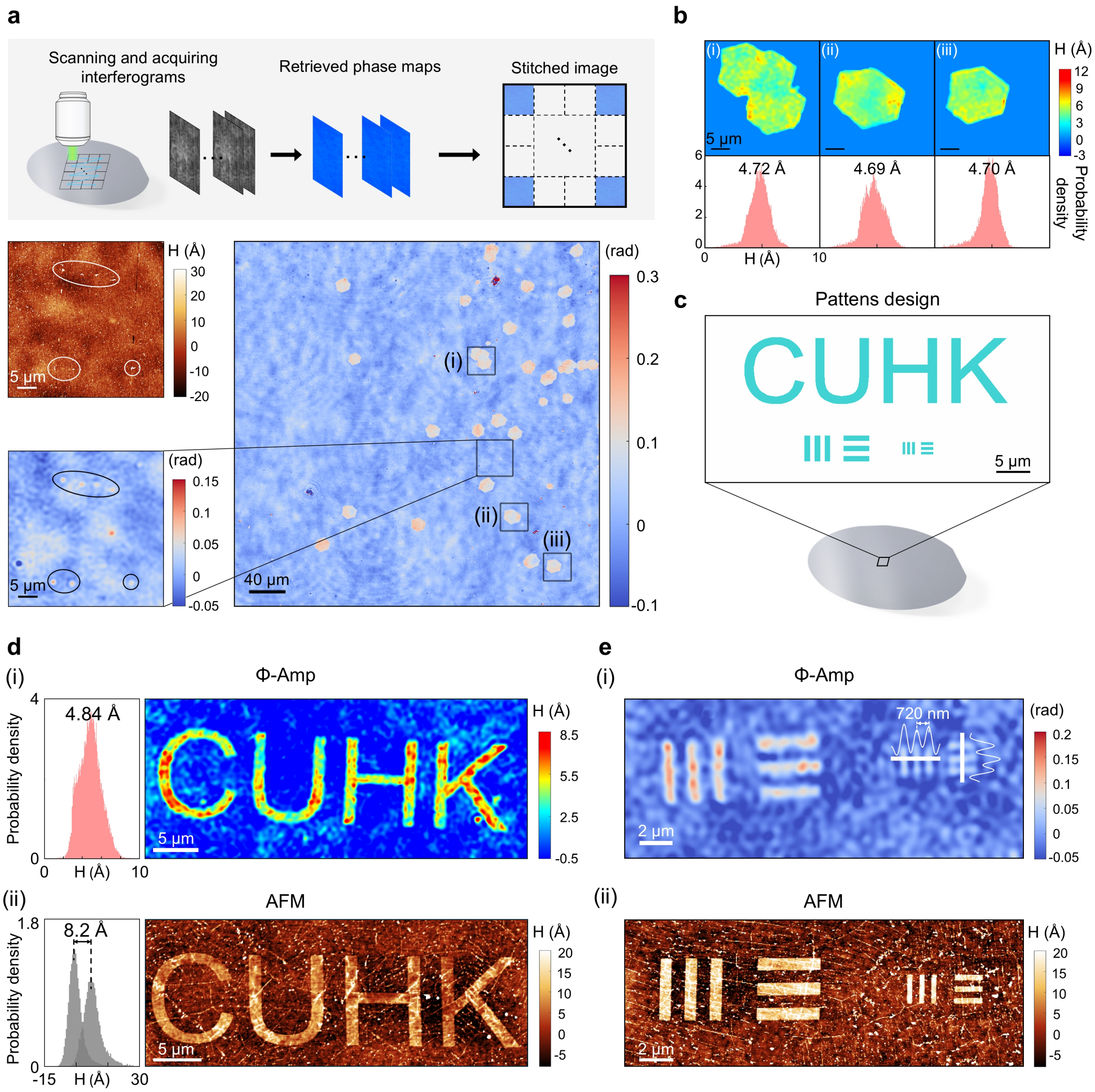

**Fig. 4. Thickness profiling performance of Φ-Amp. a,** The workflow for obtaining large FOV imaging of MLG flakes and the corresponding mapping result. We selected a region within the black box and compared the mapping result with the AFM result. The residues can be well revealed, which proves that Φ-Amp can characterize the sample quality. From this map, we further selected three MLG regions (i)-(iii) for further analysis in **b**. **b,** (i)-(iii) are the thickness maps and histograms of the corresponding selected MLG regions in **a**, from which we can calculate the thickness values of MLG flakes. **c,** The design of the CUHK logo and resolution line pairs. **d,** (i) and (ii) are the thickness maps and corresponding histograms of the CUHK logo obtained by Φ-Amp and AFM, respectively. **e**, Spatial resolution evaluation. (i) is the phase map of resolution line pairs measured by Φ-Amp, and (ii) is the thickness maps of resolution line pairs measured by AFM.

**Quantification of twistronics-structure, impurities and defects on bilayer graphene**

In twisted bilayer graphene (tBLG), variations in van der Waals and electronic interlayer interactions between atomic layers can result in an interlayer spacing difference[1, 6, 54] that reflects the electronic coupling strength, as illustrated in **Fig. 5a.** The equilibrium interlayer spacing $d_0$ is at the Å-level which can be theoretically modeled by assessing the energy convergence using self-consistent field calculations[55], while the difference of $d_0$ for different twisting configurations is at the sub-Å level. To demonstrate the capability of Φ-Amp for quantifying interlayer spacing difference in twistronics, we first synthesized tBLG samples of two stacking orders, i.e., AB-BLG and 30°-tBLG, by chemical vapor deposition (CVD) and transferred them to the optimized single-layer $SiO_2$/Si phase cavity with $SiO_2$ layer thickness of 285.8 nm. We then retrieved their phase maps and converted them into thickness maps using the reconstruction model in **Fig. 5b**, where the selected area electron diffraction (SAED) patterns of AB-BLG and 30°-tBLG are shown in the insets. In the thickness maps, we segmented the MLG and bilayer regions and plotted their histograms as shown in **Fig. 5b**. By referencing the same MLG structure, we determined the interlayer spacings of AB-BLG and 30°-tBLG of 3.75 Å, and 4.48 Å, respectively, and their differences of 0.73 Å, which is qualitatively consistent to the low-energy electron microscopy (LEEM) measurements[6]. We further repeated the measurements using three independent samples for each stacking order and conducted the statistical analysis of the interlayer spacing difference. The 95% confidence interval for the interlayer spacing difference is 0.77±0.50 Å. Details about the statistical analysis, DFT calculation and Raman spectra characterization are included in **Supplementary Note 9**.

We further characterized a bilayer graphene on the double-layer cavity by bright field microscopy, AFM, and our Φ-Amp as shown in **Fig. 5c**. The bright field imaging in **Fig. 5c(i)** cannot distinguish the bilayer graphene from monolayer graphene, as the double-layer cavity substrate is designed to enhance the phase sensitivity by squeezing the intensity sensitivity. **Figure. 5c(ii)** and **(iii)** compare Φ-Amp imaging results with AFM, the green lines indicate that some impurity features appear on both imaging modalities. Φ-Amp zoom-in plot **Fig. 5c(iv)** points out the existence of an atomic crack defect which is verified by a fine-step scanning AFM in **Fig. 5c(v)**. More results about the wavelength dependence of the BLG sample on the double-layer cavity are included in **Supplementary Note 5**.

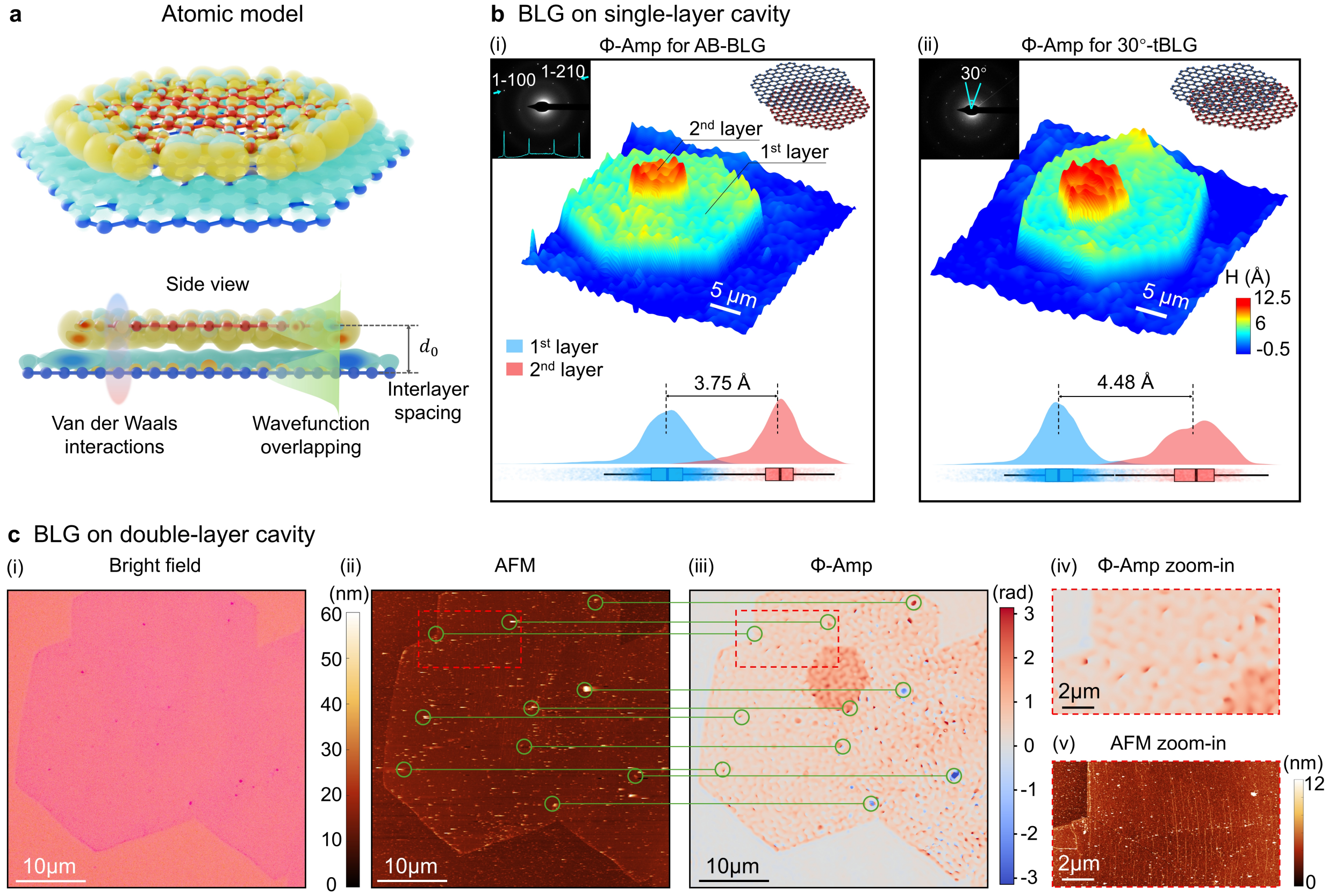

**Fig. 5. Quantification of twistronics-structure, impurities and defects on bilayer graphene. a,** Illustration of the interlayer spacing induced by the interactions between adjacent atomic layers. **b,** The SAED patterns, retrieved thickness maps, and the corresponding thickness histograms from Φ-Amp of AB-BLG and 30°-tBLG on the single-layer cavity, respectively. Scale bar: 5 μm.

**c,** Characterization of BLG on the double-layer cavity. (i) bright field image; (ii) AFM results with a step size of 117.2 nm; (iii) Φ-Amp image; (iv) zoom-in plot of Φ-Amp imaging marked by the dashed red box in (iii); (v) AFM with a fine step size of 29.3 nm. Scale of (i)-(iii) and (iv)-(v) are aligned with a scale bar of 10 μm and 2 μm, respectively. All green lines have the same length and indicate common impurity features in AFM and Φ-Amp.

## DISCUSSION

Phase Amplification microscopy is a conceptually distinct approach that bypasses the noise-suppression paradigm. By operating near the phase resonance condition, we magnify sub $10^{-5}$ rad phase shifts, far below the noise floor of conventional quantitative phase microscopy to a detectable level in a single shot, without amplifying the underlying noise. This principle allowed us to demonstrate, under ambient conditions, a thickness measurement accuracy at femtometer-level, more than two orders of magnitude beyond the current state of the art. Importantly, while the resonant cavity also enhances intensity contrast, we note that this occurs at the expense of confining light within the substrate, rendering the intensity signal largely insensitive to the 2D material layer. Our off-axis interferometric detection circumvents this limitation by directly recovering the amplified phase, providing a robust and quantitative readout of the atomic structure.

The concept of phase amplification is, in our view, only in its infancy. Just as LIGO has pushed displacement sensitivity to $10^{-19}$ m level by exploiting optical resonance in kilometer-scale cavities [43], the accuracy of Φ-Amp can be further scaled toward sub-attometer precision through the engineering of higher- quality phase cavities. This trajectory, however, demands stringent system-level improvements: (i) sub-nanometer fabrication precision and layer flatness control of the phase cavity; (ii) active stabilization of laser power and wavelength drifts to $< 10^{-16}$ m level, which needs feedback mechanisms; (iii) mechanical stability and precise temperature control; and (iv) enhanced detection dynamic range, achievable through photomultiplier-tube arrays, spatial filtering, or balanced detection. Nevertheless, these are engineering challenges—not fundamental physical barriers. The simplicity and robustness of single-layer phase cavities already offer a compelling entry point, benefiting from ease of fabrication, compatibility with existing silicon platforms, and a generous tolerance to illumination bandwidth. With a high phase SNR, one may also improve the spatial resolution in Φ-Amp via synthetic aperture illumination[56], shorter wavelength sources, and computational approaches with *a priori* information[57]. Moreover, the wavelength- scanning capability inherent to our approach provides a self- referencing route to estimate local substrate thickness, adding a valuable diagnostic dimension.

Beyond metrology, Φ- Amp opens several promising avenues. In the realm of quantum materials, the phase- gain concept is inherently generalizable: it can be integrated into existing laser- interferometric imaging modalities and extended to other 2D materials, including transition metal dichalcogenides, by tailoring the cavity parameters to their refractive indices and thicknesses. The wide-field, high-throughput capability of Φ-Amp could accelerate the development of high-performance atomic devices by enabling rapid, non-destructive quality screening of large-area fabricated structures[58]. More ambitiously, we envision real-time imaging of dynamic phenomena at the nanoscale—such as confined charge transport and self-assembly in 2D nanochannels[59, 60]. This capability may also extend to high-contrast, real-time, label-free imaging of biological processes at the sub-nanometer scale: for example, monitoring neuron action potentials, which involve minute optical path changes[34]. For decoupling small phase fluctuations from the static phase background in biological structures, the adaptive dynamic range shift (ADRIFT) phase imaging[39] can isolate the signal of interest.

In essence, we have reported Φ-Amp, which leverages the native substrate as a phase cavity to physically amplify weak phase signals before detection. This paradigm shift bridges the gap between the exquisite sensitivity of large-scale interferometry and the practicality of optical microscopy, offering a scalable platform for probing the subtle structural and electronic dynamics that define the next generation of quantum materials and devices.

**Data Availability**

The datasets generated during the current study are available from the corresponding author upon reasonable request.

**Code Availability**

The code used for simulation and data plotting is available from the corresponding author upon reasonable request.

**Author Contributions**

Conceptualization: R. Z. and N. Z. Methodology: N. Z., Q. Zhang and R. Z. Material fabrication and characterization: T. H., J. Y., J. Z., N. Z., Q. Zhang and C. X. Supervision: R. Z. and N. X. F. Computer simulations of DFT calculations: H. Y. Li and N. X. F. Manuscript writing: N. Z., Q. Zhang, N. X. F. and R. Z. Code: N. Z., Q. Zhang, C. X. and Y. N. Review & Editing: N. Z., Q. Zhang, C. X., R. Z., N. X. F., and all co-authors.

**Acknowledgments**

R. Z. acknowledges the financial support from Hong Kong Innovation and Technology Fund (Grant Nos. ITS/229/23FP and GHP/171/22SZ), Research Grant Council of Hong Kong SAR (Grant No. R4024-23, N_CUHK431/23, RFS2526-4S05 and 14202825), and NSFC Excellent Young Scientists Fund (No. 62422513); Z. G. acknowledges the financial support from the Key-Area Research and Development Program of Guangdong Province (Grant No. 2020B0101030002), the National Natural Science Foundation of China (Grant No. 62101475), the Research Grant Council of Hong Kong (Grant Nos. 24201020 and 14207421), and the Research Matching Grant Scheme of Hong Kong Government (Grant No. 8601547). J. X. acknowledges the financial support from the Research Grants Council of Hong Kong SAR (Grant No. N_CUHK 438/18) and CUHK Group Research Scheme (Grant No. AoE/P-701/20, 14203018). Q. Zhan acknowledges the visiting faculty program (2024-2025) of CUHK. R. Z., N. X. F., and J. Z acknowledge the support by the Ministry of Science and Technologies under the Grant 2023YFF1500900. N. X. F. acknowledges the Research Grants Council of Hong Kong under Grant STG 3/E-704/23-N, and the Guangdong Basic and Applied Basic Research Foundation under the grant GDZX2304003. The research work by the Fang group at HKU is conducted in the JC STEM Lab of Scalable and Sustainable Photonic Manufacturing, funded by The Hong Kong Jockey Club Charities Trust. N. X. F. also wants to thank for startup funding from Materials Innovation Institute for Life Sciences and Energy (MILES), HKU-SIRI in Shenzhen for supporting this manuscript.

**Conflict of Interest**

A US Non-provisional Patent and a China Invention Patent have been filed based on this work through The Chinese University of Hong Kong. R. Z. is a co-founder of Bay Jay Ray Technology Limited which did not fund this work.

# Supplementary Information for
# Phase amplification microscopy with femtometer-level accuracy

Nansen Zhou[1,†], Qihang Zhang[1,†], Ting Huang[1], Helios Y. Li[2], Chenjia Xi[1], Jiawen You[1,3], Jinsong Zhang[4], Yujie Nie[1], Chaoran Huang[5], Zhaoli Gao[1], Jinlong Zhu[4], Qiwen Zhan[1,6], Jianbin Xu[5], Nicholas X. Fang[2,7,8,*], and Renjie Zhou[1,5,*]

[1]Department of Biomedical Engineering, The Chinese University of Hong Kong, Shatin, New Territories, Hong Kong SAR, China

[2]Department of Mechanical Engineering, The University of Hong Kong, Hong Kong SAR, China

[3]Department of Chemical and Biological Engineering, The Hong Kong University of Science and Technology, Clear Water Bay, Kowloon, Hong Kong SAR, China

[4]State Key Laboratory of Digital Manufacturing Equipment and Technology, Huazhong University of Science and Technology, Wuhan 430074, China

[5]Department of Electronic Engineering, The Chinese University of Hong Kong, Shatin, New Territories, Hong Kong SAR, China

[6]Zhejiang Key Laboratory of 3D Micro/Nano Fabrication and Characterization, Department of Electronic and Information Engineering, School of Engineering, Westlake University, Hangzhou, Zhejiang, China

[7]HK Institute of Quantum Science & Technology, The University of Hong Kong, Pokfulam Road, Hong Kong SAR, China

[8]Materials Innovation Institute for Life Sciences and Energy (MILES), HKU-SIRI, Shenzhen, China

*Corresponding authors: nicxfang@hku.hk; rjzhou@cuhk.edu.hk

[†]These authors contributed equally to this work.

**This file includes:**

Methods

Supplementary Note 1 to Note 10

Figs. S1 to S19

References

## Extended Figures

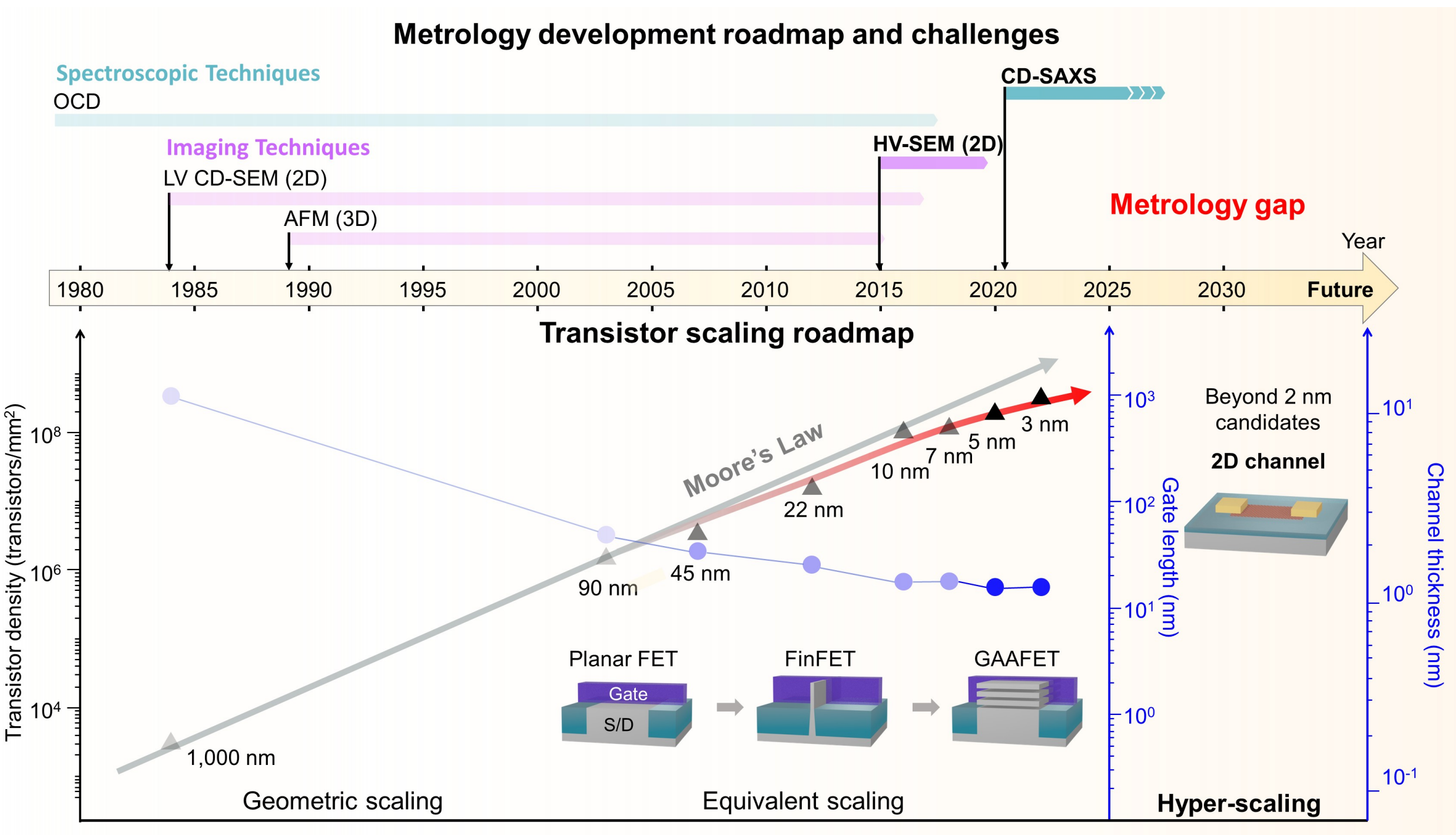

**Supplementary Fig. S1. Transistors scaling and metrology development roadmaps and grand challenges.** OCD: Optical Critical Dimension; SAXS: Small Angle X-ray Scattering; LV CD-SEM: Low Voltage Critical Dimension-SEM Scanning Electron Microscopy; HV CD-SEM: High Voltage Critical Dimension-Scanning Electron Microscopy; AFM: atomic force microscopy; CMOS: Complementary Metal-Oxide-Semiconductor. Ref. 2023 International Roadmap for Devices and Systems.

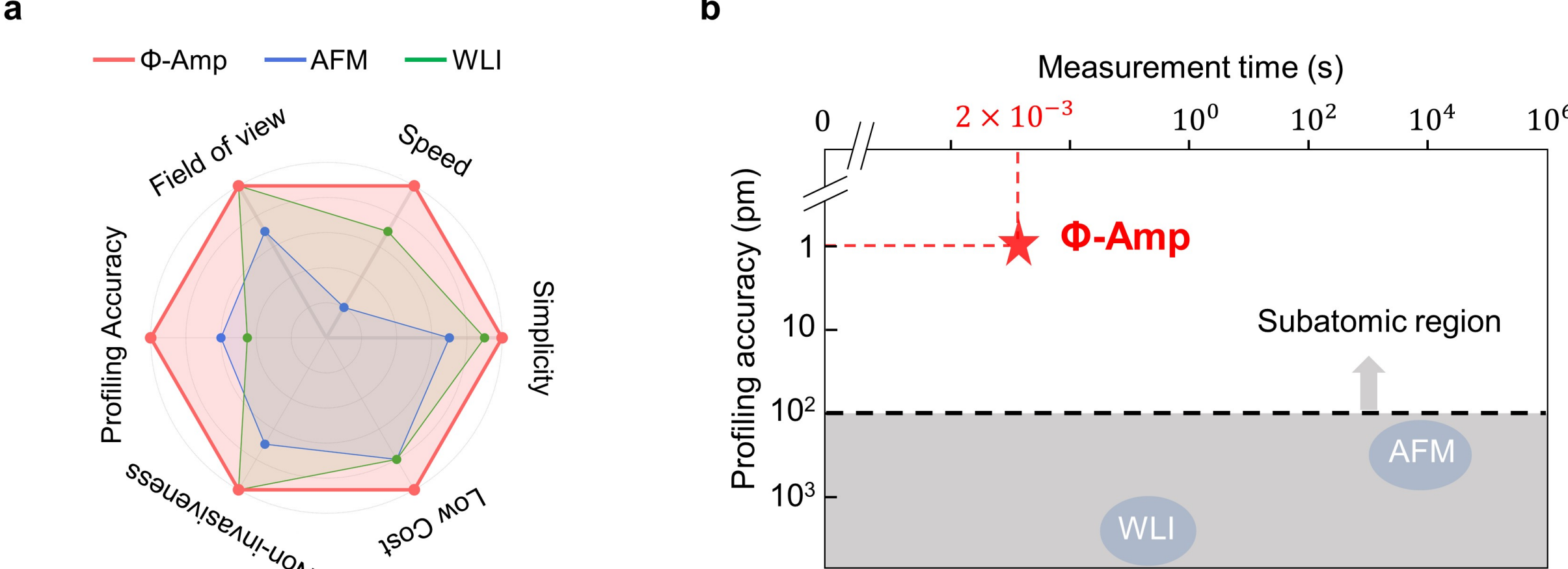

**Supplementary Fig. S2: Comparison of Φ-Amp, AFM, and WLI**. **a,** Comparison between Φ-Amp, AFM, and WLI in perspective of simplicity, speed, profiling accuracy, cost, field of view, and noninvasiveness. **b,** By comparing measurement time and accuracy, Φ-Amp achieves high throughput and 1 pm measurement accuracy.

## Methods

### Imaging system of Φ-Amp

The imaging system is shown in **Fig. 2b**[1]. A collimated laser beam with linear polarization passes through a polarization beam splitter (PBS). Then, a half-wave plate (HWP) and a quarter-wave plate (QWP) are placed after the PBS to adjust the polarization of the light wave from linear polarization to circular polarization. The light wave is then directed onto the sample at normal incidence through an objective lens. This same objective lens is also responsible for collecting the reflected beam. The QWP further converts the polarization of the reflected light into a direction perpendicular to the incident light, effectively separating the incident and reflected beams as the light passes through the PBS again. Following this, a common-path interferometer composed of a diffraction grating (DG), a pinhole, and a *4f* system, is employed to extract the phase map of the sample.

### Sample preparation

The graphene was synthesized by the atmospheric pressure chemical vapor deposition (APCVD) system in a 1-inch furnace (Lindberg Blue M™, Thermo Fisher Scientific Inc.). In brief, the copper foil (Alfa Aesar #46986, Thermo Fisher Scientific Inc.) was cut into the desired size and cleaned by sonicating in $HNO_3$ (5.4%) for 1 min, followed by rinsing with DI water twice. The copper foil was then placed at the center of the furnace and heated from room temperature to 1050 °C over 30 minutes under a mixed gas flow of 500 sccm Ar and 30 sccm $H_2$. Subsequently, 5 sccm $CH_4$ (5% diluted in Ar) was injected into the system for 1h while the temperature was held constant. The furnace was moved away from the copper foil for fast cooling.

The as-grown bilayer graphene was then transferred to a 285.8 nm $SiO_2$/Si substrate using the bubbling transfer method. First, the Graphene/Cu was spin-coated with PMMA (950 PMMA A4, MicroChem

Co.) at 4000 rpm for 1 min, followed by baking at 105°C for 2 min. The PMMA/Gr/Cu stack was connected to the anode and immersed in 50 mM NaOH aqueous solution, with a platinum wire connected to the cathode of the power supply. Upon applying ~ 10 V voltage, hydrogen bubbles generated at the interface gently lifted the PMMA/Graphene off the copper substrate. The PMMA/Graphene stack was then rinsed with a deionized (DI) water bath to remove residual ions. Subsequently, the PMMA/Graphene film was transferred onto the $SiO_2$/Si wafer and air-dried for 2 hours. The PMMA layer was then dissolved in acetone. Finally, the transferred graphene was annealed under a mixture of Ar and H2 atmosphere in a vacuum condition at 275°C for 1h.

To pattern the graphene, the as-grown monolayer graphene film was first transferred onto a 285.8 nm $SiO_2$/Si wafer by the abovementioned process and then spin-coated with PMMA (950 PMMA A3, MicroChem Corp.) as e-beam resist. The desired patterns were created by electron beam lithography (ELPHY Plus, Raith GmbH). After developing in MIBK: isopropanol (3:1 in volume ratio) solution, the exposed graphene was etched by RIE (Oxford Instruments plc) with a 100 sccm $O_2$ to form the pattern, and then the PMMA can be removed by acetone.

**Repeatability, sensitivity, accuracy, and spatial resolution**

Repeatability describes an instrument's ability to yield consistent results under identical conditions, governed primarily by system stability. Sensitivity refers to the smallest detectable change in a measured quantity, often limited by noise sources such as mechanical vibrations, air disturbances, photon shot noise, and detector noise. In a high-sensitivity system, high repeatability is needed to guarantee a reliable detection of the smallest detectable signal rather than noise. Accuracy, distinct from both, quantifies how closely measurements align with a reference value, and it is sample dependent; achieving high accuracy demands not only repeatability and sensitivity but also a precise sample reconstruction model to minimize systematic errors. Spatial resolution defines the smallest distinguishable separation between two points by the system, which is influenced by the sensitivity and signal, i.e., SNR. Repeatability, sensitivity, accuracy, and spatial resolution collectively define the performance of a measurement system[2].

**Sample characterization**

AFM imaging was performed by Bruker MultiMode 8-HR at a 0.5 Hz scanning rate, equipped with a probe (RTESPA-300, Bruker Corp.). The quantitative analysis of the image profiles was carried out using Gwyddion software. Raman spectra were measured on a RENISHAW inVia™ confocal Raman microscope with an excitation laser wavelength of 532 nm. The thicknesses of $SiO_2$ and $Si_3N_4$ films were measured by a J.A. Woollam RC2 XI ellipsometer with a spectral range of 210-2500 nm at incident angles of 65° and 75°.

**Measurement conditions of Φ-Amp**

The Φ-Amp measurements were conducted in a stable and controlled environment. The ambient temperature was kept constant within a specific range (20-22°C) to prevent thermal expansion or contraction of the sample. The humidity level was controlled (30-50% relative humidity) to avoid condensation and sample oxidation. An air-floating vibration isolation optical table (VIH150300-24, VERE Inc.) was used to alleviate external mechanical vibrations. The ambient light during measurement was minimized to prevent optical distortions. Additionally, the central wavelength of the monochromatic laser is 532 nm (MGL-FN-532-400mW, Changchun New Industries Optoelectronics Technology Co., Ltd.), and the FWHM is 0.3 nm by measuring the spectrum using a spectrometer (HR4000, Ocean Optics Inc.). We used a high electron-well-capacity camera with a full electron-well-capacity of $2\times10^6$ electrons (Q-2HFW-Hm/CXP-6, Adimec Advanced Image

Systems B.V.) to capture the interferograms. It should be noted that Φ-Amp can also achieve quantitative mapping of monolayer graphene using a regular camera with a full electron-well capacity of $1\times10^4$ electrons, but the retrieved thickness map has a lower CNR value than the results obtained by a high well-capacity camera. The detailed comparison is shown in **Supplementary Note 10**.

**Supplementary Note 1. Phase gain theory**

In Section 1.A, we propose a transfer-matrix model for modeling light propagation inside a multilayer phase cavity. Using the model, we derive the amplified phase and original phase of a single-layer phase cavity case, and we define the phase gain for a single-layer phase cavity in Section 1.B. In Section 1.C, we generalize our phase gain theory to a multilayer phase cavity condition.

**A. Modeling of light propagation in multilayer phase cavity**

The multilayer phase cavity is modeled as a sample layer, *N-2* thin film layers, and a substrate (denoted as the $N^{th}$ layer), and the complex refractive index value of each layer is $\tilde{n}_j$, $j = 1,2,\ldots N$, surrounded by a uniform medium ($\tilde{n}_0$), as illustrated in **Fig. S4**. The incident field $E_i$, reflected field $E_r$, and the transmitted field $E_t$ can be connected by the transfer-matrix $\mathrm{M}$ as below [3, 4]:

$$\begin{bmatrix} E_i \\ E_r \end{bmatrix} = \mathrm{M} \begin{bmatrix} E_t \\ 0 \end{bmatrix} = \begin{bmatrix} m_{11} & m_{12} \\ m_{21} & m_{22} \end{bmatrix} \begin{bmatrix} E_t \\ 0 \end{bmatrix}, \tag{S1}$$

where $\mathrm{M} = \mathrm{D}_{0,1} \cdots \mathrm{D}_{j-1,j} \mathrm{P}_j \mathrm{D}_{j,j+1} \mathrm{P}_{j+1} \mathrm{D}_{j+1,j+2} \cdots \mathrm{D}_{\mathrm{N-1,N}} \mathrm{P}_{\mathrm{N}} \mathrm{D}_{\mathrm{N,N+1}}$, $\mathrm{D}$ and $\mathrm{P}$ represent the transmission matrix at each interface and the propagation matrix in each layer, respectively. In the transmission matrix $\mathrm{D}_{j,j+1}$, the subscripts *j* and *j*+1 denote the indices of two adjacent layers forming the interface. The propagation matrix $\mathrm{P}_j$ corresponds to light traveling through layer $j$. Using Eq. (S1), the reflected field $E_r$ can be directly related to the incident field expressed by the matrix component as $E_r = \dfrac{m_{21}}{m_{11}} E_i$.

As illustrated in **Fig. S4**, the transmitted field $E_{t,j}^{j,j+1}$ and reflected field $E_{r,j}^{j,j+1}$ in layer *j* at the interface between layer *j* and *j*+1 are related by the reflection coefficient $r_{j,j+1}$ and transmission coefficients $t_{j,j+1}$:

$$E_{t,j+1}^{j,j+1} = t_{j,j+1} E_{t,j}^{j,j+1} + r_{j+1,j} E_{r,j+1}^{j,j+1}, \tag{S2a}$$

$$E_{r,j}^{j,j+1} = r_{j,j+1} E_{t,j}^{j,j+1} + t_{j+1,j} E_{r,j+1}^{j,j+1}, \tag{S2b}$$

where $r_{j,j+1}$ and $t_{j,j+1}$ can be determined with the complex refractive index values of layer *j* and layer *j*+1:

$$r_{j,j+1} = \frac{\tilde{n}_j - \tilde{n}_{j+1}}{\tilde{n}_j + \tilde{n}_{j+1}}; \quad t_{j,j+1} = \frac{2\tilde{n}_j}{\tilde{n}_j + \tilde{n}_{j+1}}, \tag{S3}$$

and

$$r_{j,j+1} = -r_{j+1,1}; \quad t_{j,j+1} t_{j+1,j} - r_{j,j+1} r_{j+1,j} = 1. \tag{S4}$$

According to Eq. (S4), we can rewrite Eq. (S2) in matrix form and define the transmission matrix $\mathrm{D}_{j,j+1}$:

$$\begin{bmatrix} E_{t,j}^{j,j+1} \\ E_{r,j}^{j,j+1} \end{bmatrix} = \frac{1}{t_{j,j+1}} \begin{bmatrix} 1 & r_{j,j+1} \\ r_{j,j+1} & 1 \end{bmatrix} \begin{bmatrix} E_{t,j+1}^{j,j+1} \\ E_{r,j+1}^{j,j+1} \end{bmatrix} = \mathrm{D}_{j,j+1} \begin{bmatrix} E_{t,j+1}^{j,j+1} \\ E_{r,j+1}^{j,j+1} \end{bmatrix}, \tag{S5}$$

Assuming a plane wave at normal incidence, the transmitted and reflected fields inside the layer *j* can be connected as:

$$E_{t,j}^{j,j+1}(z) = E_{t,j}^{j-1,j}(z + H_j), \tag{S6a}$$

$$E_{r,j}^{j-1,j}(z) = E_{r,j}^{j,j+1}(z - H_j). \tag{S6b}$$

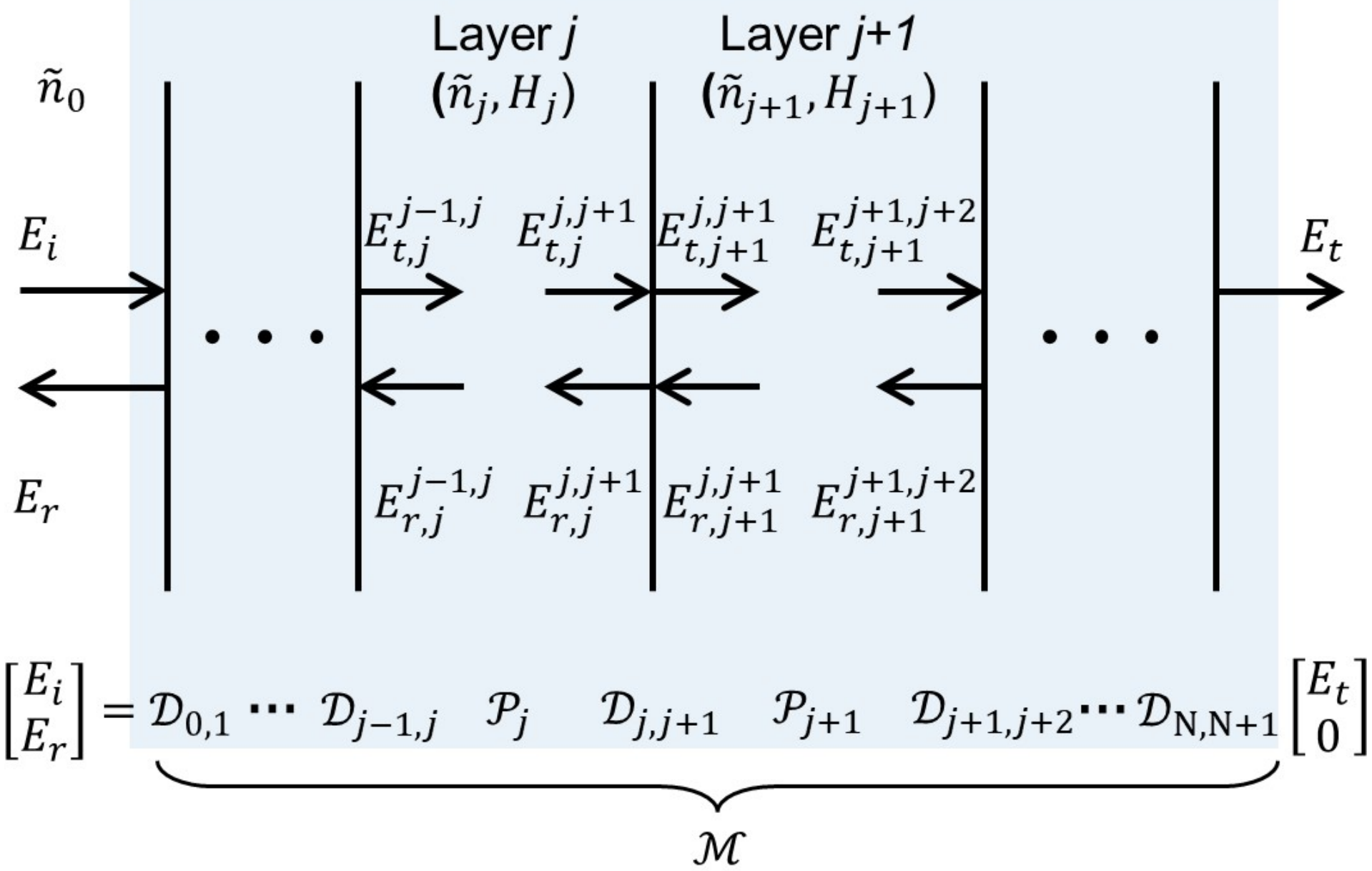

**Supplementary Fig. S4. Illustration of the transfer-matrix model for modeling of light propagation in a multilayer specimen.** $E_i$, $E_r$ and $E_t$ represent the incident field, the reflected field, and the transmitted field, respectively; $\tilde{n}_j$ represents the complex refractive index of the material in layer $j$; $H_j$ represents the thickness of layer $j$; $E_{r,j}^{j,j+1}$ and $E_{t,j}^{j,j+1}$ present the reflected and transmitted fields in the layer $j$ at the interface between layer $j$ and layer $j$+1.

We can rewrite Eq. (S6) in matrix form and define the propagation matrix $\mathrm{P}_j$ inside layer $j$:

$$\begin{bmatrix} E_{t,j}^{j-1,j} \\ E_{r,j}^{j-1,j} \end{bmatrix} = \begin{bmatrix} e^{-ik_j H_j} & 0 \\ 0 & e^{ik_j H_j} \end{bmatrix} \begin{bmatrix} E_{t,j}^{j,j+1} \\ E_{r,j}^{j,j+1} \end{bmatrix} = \mathrm{P}_j \begin{bmatrix} E_{t,j}^{j,j+1} \\ E_{r,j}^{j,j+1} \end{bmatrix}, \tag{S7}$$

where $k_j = 2\pi \tilde{n}_j / \lambda$ with the incident wavelength $\lambda$ in free space.

In this work, we will use silicon with a thickness of hundreds of microns as the substrate and visible-light illumination. The light will be attenuated within the substrate due to the non-negligible imaginary part of the complex refractive index of silicon, resulting in $\left|e^{2jk_N H_N}\right| \cong 0$. Therefore, $E_r$ can be expressed as:

$$E_r = \frac{m_{21}^{'} + m_{22}^{'} r_{N,N+1} e^{2jk_N H_N}}{m_{11}^{'} + m_{12}^{'} r_{N,N+1} e^{2jk_N H_N}} E_i \cong \frac{m_{21}^{'}}{m_{11}^{'}} E_i. \tag{S8}$$

where $\mathrm{D_{0,1}P_1D_{1,2}P_2 \cdots D_{N-1,N}} = \begin{bmatrix} m_{11}^{'} & m_{12}^{'} \\ m_{21}^{'} & m_{22}^{'} \end{bmatrix}$. Eq. (S8) indicates that the propagation matrix $\mathrm{P_N}$ inside opaque substrate and the transmission matrix $\mathrm{D}_{N,N+1}$ at the interface between the substrate and surrounding medium do not contribute to the reflected field $E_r$.

**B. Phase gain *G* derivation for a single-layer phase cavity**

We first consider a single-layer phase cavity including the sample layer ($\tilde{n}_1$), cavity layer ($\tilde{n}_2$), and the silicon substrate ($\tilde{n}_3$), as illustrated in **Fig. S5a.** Using Eq. (S8), the total reflected wave can be expressed as,

$$E_r = \frac{e^{i2\varphi_1}\left(r_{1,2} + r_{2,3} e^{i2\varphi_2}\right) + r_{0,1}\left(1 + r_{1,2} r_{2,3} e^{i2\varphi_2}\right)}{r_{0,1} e^{i2\varphi_1}\left(r_{1,2} + r_{2,3} e^{i2\varphi_2}\right) + \left(1 + r_{1,2} r_{2,3} e^{i2\varphi_2}\right)}, \tag{S9}$$

where $\varphi_j = k_j H_j \left(j = 1,2\right)$. The formula of Eq.(S9) is complex, and its complexity increases with the number of layers. To simplify the expression in Eq.(S9), we introduced the equivalent reflection interface and equivalent reflection coefficient. In such a *3*-layer system, the light propagation from layer *2* to layer *3* can be equivalent to the reflection at an equivalent reflection interface with an equivalent reflection coefficient of $r_{1,3}$, which depends on the reflection coefficient $r_{2,3}$ of the interface between layer *2* and layer *3*, and the reflection coefficient $r_{1,2}$ of the interface between layer *1* and layer *2* (**Fig. S5b(i)**), i.e., $r_{1,3}=\dfrac{r_{1,2}+r_{2,3}e^{i2\varphi_2}}{1+r_{1,2}r_{2,3}e^{i2\varphi_2}}$. Therefore, Eq. (S9) can be rewritten as:

$$E_r=\frac{r_{0,1}+r_{1,3}e^{i2\varphi_1}}{1+r_{0,1}r_{1,3}e^{i2\varphi_1}}, \tag{S10}$$

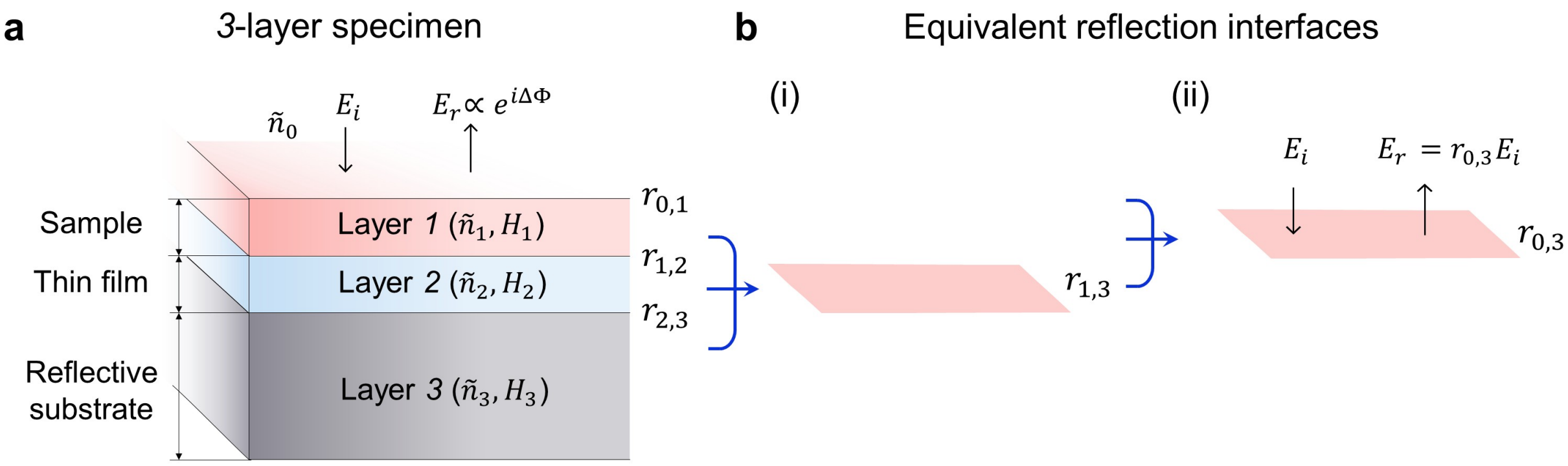

**Supplementary Fig. S5. Simplification of light propagation in a *3*-layer specimen to the reflected light wave at the equivalent reflection interface**.

Then, the light propagation from layer *1* to layer *3* can be equivalent to the reflection at an equivalent reflection interface with an equivalent reflection coefficient $r_{0,3}$, which depends on the equivalent reflection coefficient $r_{1,3}$ and the reflection coefficient $r_{0,1}$ of the interface between the surrounding medium and layer *1* (**Fig. S5b(ii)**). Therefore, Eq. (S10) can be written as $E_r=|E_r|e^{i\Phi}=r_{0,3}$.

Analogy to the definition of reflection coefficient by refractive index, $r_{0,3}$ can be defined by effective complex refractive index values $\tilde{n}^{(3)}_{0,eff}$ and $\tilde{n}^{(3)}_{3,eff}$ as $r_{0,3}=\dfrac{\tilde{n}^{(3)}_{0,eff}-\tilde{n}^{(3)}_{3,eff}}{\tilde{n}^{(3)}_{0,eff}+\tilde{n}^{(3)}_{3,eff}}$. Based on the equivalence of light propagation inside a single-layer phase cavity system, we can derive the recursive relationship between the equivalent refractive index values:

$$\tilde{n}_{0,eff}^{(3)} = n_0 \left( \tilde{n}_{1,eff}^{(2)} cos\varphi_1 - i\tilde{n}_{3,eff}^{(2)} sin\varphi_1 \right), \tag{S11-a}$$

$$\tilde{n}_{3,eff}^{(3)} = \tilde{n}_1 \left( \tilde{n}_{3,eff}^{(2)} cos\varphi_1 - i\tilde{n}_{1,eff}^{(2)} sin\varphi_1 \right), \tag{S11-b}$$

and

$$\tilde{n}_{1,eff}^{(2)} = \tilde{n}_1 \left( \tilde{n}_2 cos\varphi_2 - i\tilde{n}_3 sin\varphi_2 \right), \tag{S12-a}$$

$$\tilde{n}_{3,eff}^{(2)} = \tilde{n}_2 \left( \tilde{n}_3 cos\varphi_2 - i\tilde{n}_2 sin\varphi_2 \right). \tag{S12-b}$$

where $r_{1,3} = \dfrac{\tilde{n}_{1,eff}^{(2)} - \tilde{n}_{3,eff}^{(2)}}{\tilde{n}_{1,eff}^{(2)} + \tilde{n}_{3,eff}^{(2)}}$ .

In analogy to $E_r$, the reflected field at background region is $E_{r,bg} = r_{0,3,bg} = \dfrac{\tilde{n}_{0,eff,bg}^{(3)} - \tilde{n}_{3,eff,bg}^{(3)}}{\tilde{n}_{0,eff,bg}^{(3)} + \tilde{n}_{3,eff,bg}^{(3)}}$, where $\tilde{n}_{0,eff,bg}^{(3)}$ and $\tilde{n}_{3,eff,bg}^{(3)}$ can be obtained by setting $\tilde{n}_1 = n_0$ in Eq. (S11),

$$\tilde{n}_{0,eff,bg}^{(3)} = n_0 \left( \tilde{n}_{1,eff,bg}^{(2)} cos\varphi_{1,bg} - i\tilde{n}_{3,eff,bg}^{(2)} sin\varphi_{1,bg} \right), \tag{S13-a}$$

$$\tilde{n}_{3,eff,bg}^{(3)} = n_0 \left( \tilde{n}_{3,eff,bg}^{(2)} cos\varphi_{1,bg} - i\tilde{n}_{1,eff,bg}^{(2)} sin\varphi_{1,bg} \right), \tag{S13-b}$$

where $\varphi_{1,bg} = k_0 H_1 = \varphi_1 n_0 / \tilde{n}_1$, $\tilde{n}_{1,eff,bg}^{(2)}$ and $\tilde{n}_{3,eff,bg}^{(2)}$ can be obtained by setting $\tilde{n}_1 = n_0$ in Eq. (S12),

$$\tilde{n}_{1,eff,bg}^{(2)} = n_0 \tilde{n}_{1,eff}^{(2)} / \tilde{n}_1, \qquad \tilde{n}_{3,eff,bg}^{(2)} = \tilde{n}_{3,eff}^{(2)}. \tag{S14}$$

By using Eq. (S13) and Eq. (S14), the reference field $E_{r,bg}$ can be rewritten as:

$$E_{r,bg} = \left|E_{r,bg}\right| e^{i\Phi_{bg}} = \frac{n_0 \tilde{n}_{1,eff}^{(2)} - \tilde{n}_1 \tilde{n}_{3,eff}^{(2)}}{n_0 \tilde{n}_{1,eff}^{(2)} + \tilde{n}_1 \tilde{n}_{3,eff}^{(2)}} e^{i2\varphi_{1,bg}}. \tag{S15}$$

Therefore, the measured phase $\Delta\Phi$ is

$$\Delta\Phi := \Phi - \Phi_{bg} = angle\left\{\frac{E_r}{E_{r,bg}}\right\}. \tag{S16}$$

For the case without a phase cavity, i.e., ($H_2 = 0$), the reflected wave $E_r^{'}$ can be written as:

$$E_r^{'} = \left|E_r^{'}\right| e^{i\varphi} = r_{0,2} = \frac{\tilde{n}_{0,eff}^{(2)} - \tilde{n}_{2,eff}^{(2)}}{\tilde{n}_{0,eff}^{(2)} + \tilde{n}_{2,eff}^{(2)}}, \tag{S17}$$

According to the recursive relation between effective complex refractive index values, we can obtain:

$$\tilde{n}_{0,eff}^{(2)} = n_0 \left(\tilde{n}_1 cos\varphi_1 - i\tilde{n}_3 sin\varphi_1\right), \tag{S18-a}$$

$$\tilde{n}_{2,eff}^{(2)} = \tilde{n}_1 \left(\tilde{n}_3 cos\varphi_1 - i\tilde{n}_1 sin\varphi_1\right). \tag{S18-b}$$

In analogy to $E_r^{'}$, the reflected field at background region is $E_{r,bg}^{'} = r_{0,2,bg} = \frac{\tilde{n}_{0,eff,bg}^{(2)} - \tilde{n}_{2,eff,bg}^{(2)}}{\tilde{n}_{0,eff,bg}^{(2)} + \tilde{n}_{2,eff,bg}^{(2)}}$, where $\tilde{n}_{0,eff,bg}^{(2)}$ and $\tilde{n}_{2,eff,bg}^{(2)}$ can be obtained by setting $\tilde{n}_1 = n_0$ in Eq. (S18):

$$\tilde{n}_{0,eff,bg}^{(2)} = n_0 \left(n_0 cos\varphi_{1,bg} - i\tilde{n}_3 sin\varphi_{1,bg}\right), \tag{S19-1}$$

$$\tilde{n}_{2,eff,bg}^{(2)} = n_0 \left(\tilde{n}_3 cos\varphi_{1,bg} - i n_0 sin\varphi_{1,bg}\right). \tag{S19-2}$$

By using Eq. (S19), the reference field $E_{r,bg}^{'}$ can be rewritten as:

$$E_{r,bg}^{'} = \left|E_{r,bg}^{'}\right| e^{i\varphi_{bg}} = \frac{n_0 - \tilde{n}_3}{n_0 + \tilde{n}_3} e^{i2\varphi_{1,bg}}. \tag{S20}$$

By using Eq. (S17) and (S20), the measured phase with the phase cavity $\Delta\varphi$ can be obtained:

$$\Delta\varphi := \varphi - \varphi_{bg} = angle\left\{\frac{E_r^{'}}{E_{r,bg}^{'}}\right\}. \tag{S21}$$

To quantitatively evaluate the phase amplification of the cavity design, we defined the phase gain $G$ using the phase measured with the phase cavity ($\Delta\Phi$) and the original phase ($\Delta\varphi$) as

$$G\left\{\tilde{n}_i; H_i; \lambda\right\} = \left|\frac{\partial(\Delta\Phi)}{\partial(H_1)} \Big/ \frac{\partial(\Delta\varphi)}{\partial(H_1)}\right|, \tag{S22}$$

where $j = 1, 2, \cdots, N$, $G$ is related to refractive index $\tilde{n}_j$, thickness $H_j$ of each layer, and the incident wavelength $\lambda$. It should be noted that we focus on imaging thin materials that require phase amplification and $\varphi_1$ is small, therefore, Eq. (S10) can be further simplified according to the Taylor series expansion approximation,

$$E_r \cong \frac{\left(n_0\tilde{n}_{1,eff}^{(2)} - \tilde{n}_1\tilde{n}_{3,eff}^{(2)}\right) + i\varphi_1\left(\tilde{n}_1\tilde{n}_{1,eff}^{(2)} - n_0\tilde{n}_{3,eff}^{(2)}\right)}{\left(n_0\tilde{n}_{1,eff}^{(2)} + \tilde{n}_1\tilde{n}_{3,eff}^{(2)}\right) - i\varphi_1\left(\tilde{n}_1\tilde{n}_{1,eff}^{(2)} + n_0\tilde{n}_{3,eff}^{(2)}\right)}, \tag{S23}$$

By combining Eq. (S15), Eq. (S16), and Eq. (S23), we can get

$$\Delta\Phi \cong \mathrm{Re}\left\{\frac{2\left(\tilde{n}_1^2 - 1\right)\left(\tilde{n}_{1,eff}^{(2)}\right)^2}{\left(\tilde{n}_{1,eff}^{(2)}\right)^2 - \left(\tilde{n}_1\tilde{n}_{3,eff}^{(2)}\right)^2}\right\}\varphi_{1,bg}, \tag{S24}$$

where the approximation $1 + ix \cong e^{ix}$ (if $x << 1$) is used since $|\varphi_1| \ll 1$, and $n_0 = 1$ for air medium, $\mathrm{R}e\{\}$ represents the real part. Similarly, we can simplify $\Delta\varphi$ as,

$$\Delta\varphi \cong \mathrm{Re}\left\{\frac{2\left(\tilde{n}_1^2-1\right)}{n_0^2-\tilde{n}_3^2}\right\}\varphi_{1,bg}\,. \quad \text{(S25)}$$

It should be noted that when the sample material matches the reflective substrate, Eq. (S25) can be reduced to $\Delta\varphi = 2\varphi_{1,bg}$, consistent with the conventional model. According to Eq. (S24) and Eq. (S25), we found that both $\Delta\Phi$ and $\Delta\varphi$ have a linear relationship with $H_1$. Therefore, the phase gain in Eq. (S22) is unrelated to $H_1$, and *G* can be also expressed as $\left|\Delta\Phi/\Delta\varphi\right|$.

**C. Generalization of *G* to a multilayer cavity**

For a multilayer cavity with an *N*-layer system, the total reflected wave expression obtained using Eq. (S8) is complex. To simplify the expression, we equate the light propagation from layer *l+1* to layer *N* to the reflection at an equivalent reflection interface with an equivalent reflection coefficient of $r_{l,N}$.

By using the recursive relationship between $r_{l,N}$ and $r_{l+1,N}$:

$$r_{l,N} = \frac{r_{l,l+1} + r_{l+1,N}e^{i2\varphi_{l+1}}}{1 + r_{l,l+1}r_{l+1,N}e^{i2\varphi_{l+1}}}\,, \quad \text{(S26)}$$

the light propagation from layer *1* to layer *N* can be equivalent to the reflection at an equivalent reflection interface with an equivalent reflection coefficient of $r_{0,N}$, i.e., $E_r = \left|E_r\right|e^{i\Phi} = r_{0,N}$. Eq. (S26) agrees with previous works for modeling light propagation in a multilayer specimen[5].

The equivalent reflection coefficient $r_{l,N}$ can be defined by effective complex refractive index values $\tilde{n}_{l,eff}^{(m)}$ and $\tilde{n}_{N,eff}^{(m)}$ for equivalent layer *l* and *N*, where *m=N-l*. According to Eq. (S26), to calculate the equivalent reflection coefficient for the *N*-layer system, the effective coefficient from layer *2* to *N* should be determined, which will in turn require the effective coefficient from layer *3* to *N*, and so forth. The recursive relationship between effective reflection coefficients $\tilde{n}_{l,eff}^{(m)}$, $\tilde{n}_{N,eff}^{(m)}$ and $\tilde{n}_{l+1,eff}^{(m-1)}$, $\tilde{n}_{N,eff}^{(m-1)}$ can be obtained using Eq. (S26),

$$\tilde{n}_{l,eff}^{(m)} = \tilde{n}_{l,eff}^{(1)} \left( \tilde{n}_{l+1,eff}^{(m-1)} cos\varphi_{l+1} - i\tilde{n}_{N,eff}^{(m-1)} sin\varphi_{l+1} \right), \quad \text{(S27-a)}$$

$$\tilde{n}_{N,eff}^{(m)} = \tilde{n}_{l+1,eff}^{(1)} \left( \tilde{n}_{N,eff}^{(m-1)} cos\varphi_{l+1} - i\tilde{n}_{l+1,eff}^{(m-1)} sin\varphi_{l+1} \right). \quad \text{(S27-b)}$$

It should be noted that the effective complex refractive index value of layer *l* will be equal to the refractive index of layer *l* for *m* =1, i.e., $\tilde{n}_{l,eff}^{(1)} = \tilde{n}_l$. By using the recursive relationship between effective reflection coefficients in Eq. (26) and Eq. (27), the measured phase with phase cavity $\Delta\Phi$ and without phase cavity $\Delta\varphi$ can be derived similarly, from which the phase gain is obtained.

## Supplementary Note 2. The optical contrast of graphene imaging

We used intensity contrast $C := \frac{|E_{r,ref}|^2 - |E_r|^2}{|E_{r,ref}|^2} = 1 - |\frac{E_r}{E_{r,ref}}|^2$ as in the literature[6] to evaluate the optical visibility of MLG. We simulated the optical contrast of MLG as a function of wavelength $\lambda$ and $SiO_2$ thickness $H_2$ as shown in **Fig. S6a.** The simulation results align with prior literature, though subtle differences exist in the wavelength-dependent variation of optical contrast[6]. This discrepancy arises from our use of wavelength-dependent refractive indices for MLG, whereas the reference[6] assumed a constant refractive index across wavelengths. Next, to experimentally verify the intensity contrast variation, we extracted the amplitude maps from the interferograms used in **Fig. 2f**. **Figure. S6b** shows the experimental verification results for the white line region in **Fig. S6a** ($\lambda$ = 532 nm, $H_2$=200 ~ 400 nm). We found that the calculated intensity contrast values in the MLG regions agree with predicted values from our phase-gain model. For the sample with a $SiO_2$ layer thickness of 260.3 nm, the contrast is around 0.156. When $H_2$ = 308.2 nm and $H_2$ = 320.3 nm, the intensity contrast is so small that it is difficult to distinguish between the sample and the background region.

To explore the effect of the imaginary part of the complex refractive index on the asymmetry of the relation between $G$ and $H_2$ in **Fig. 2**, we varied the ratio between the imaginary part and the real part of the refractive index of graphene $R = \mathrm{Im}\{\tilde{n}_1\} / \mathrm{Re}\{\tilde{n}_1\}$. We simulated the relations between $\Delta\Phi$ and $H_2$ with varying $R$ values as shown in **Fig. S7**. As $R$ changes from 0 to 1, the asymmetry becomes more pronounced. It indicates that the imaginary part of the sample arising from sample absorption mainly contributes to the asymmetrical resonance. Additionally, we found that as the absorption of the sample layer increases, the maximum value of the phase gain of the cavity will also increase.

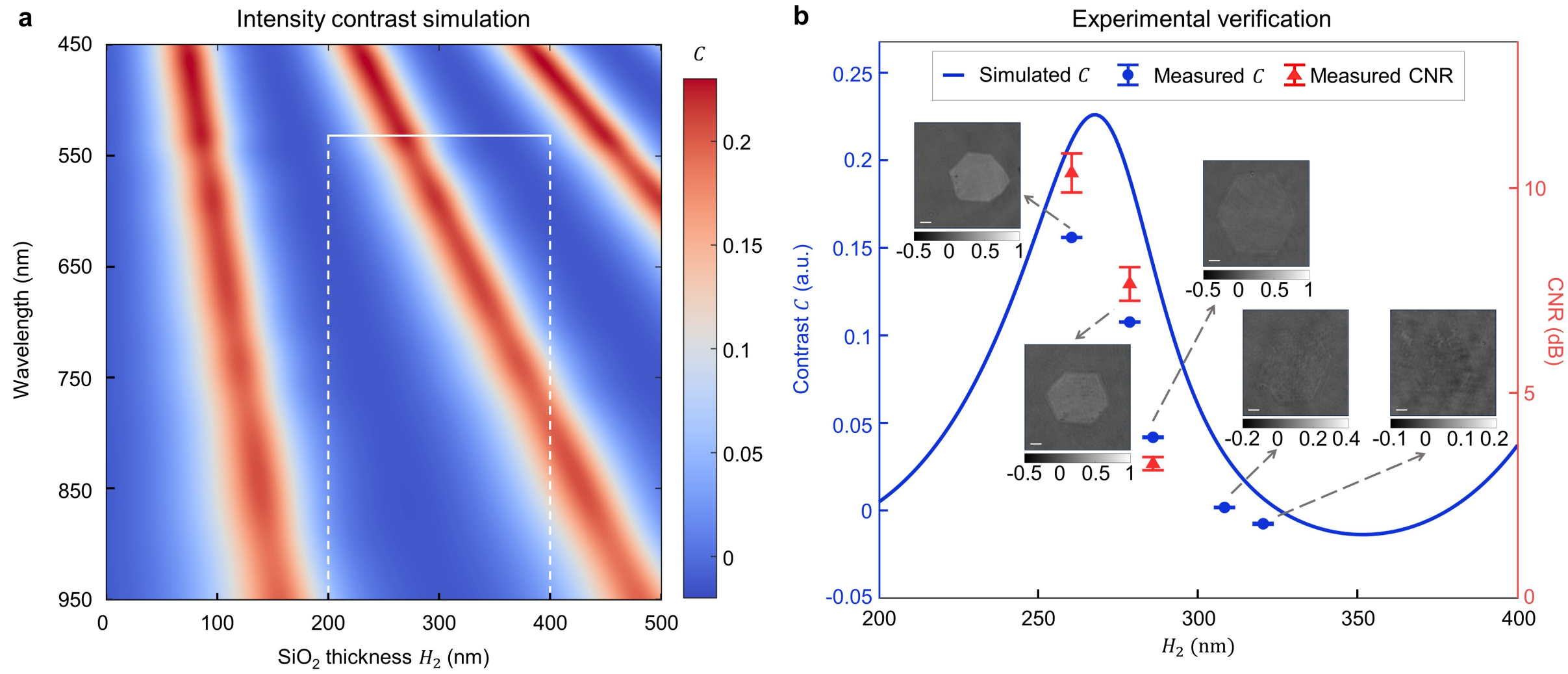

**Supplementary Fig. S6. The simulation and experimental verification of the intensity contrast of MLG. a,** The intensity contrast simulation as a function of wavelength $\lambda$ and $SiO_2$ thickness $H_2$. In the simulation, we used the calculated refractive index of graphene (refer to **Fig. 5**). **b,** Simulated $C$ v.s. $H_2$ over the white line region in **a**. When $H_2$ = 260.3 nm, 278.5 nm, 285.8 nm, 308.2 nm, and 320.3 nm, we show the measured intensity contrast maps of MLG samples. Scale bar: 5 μm.

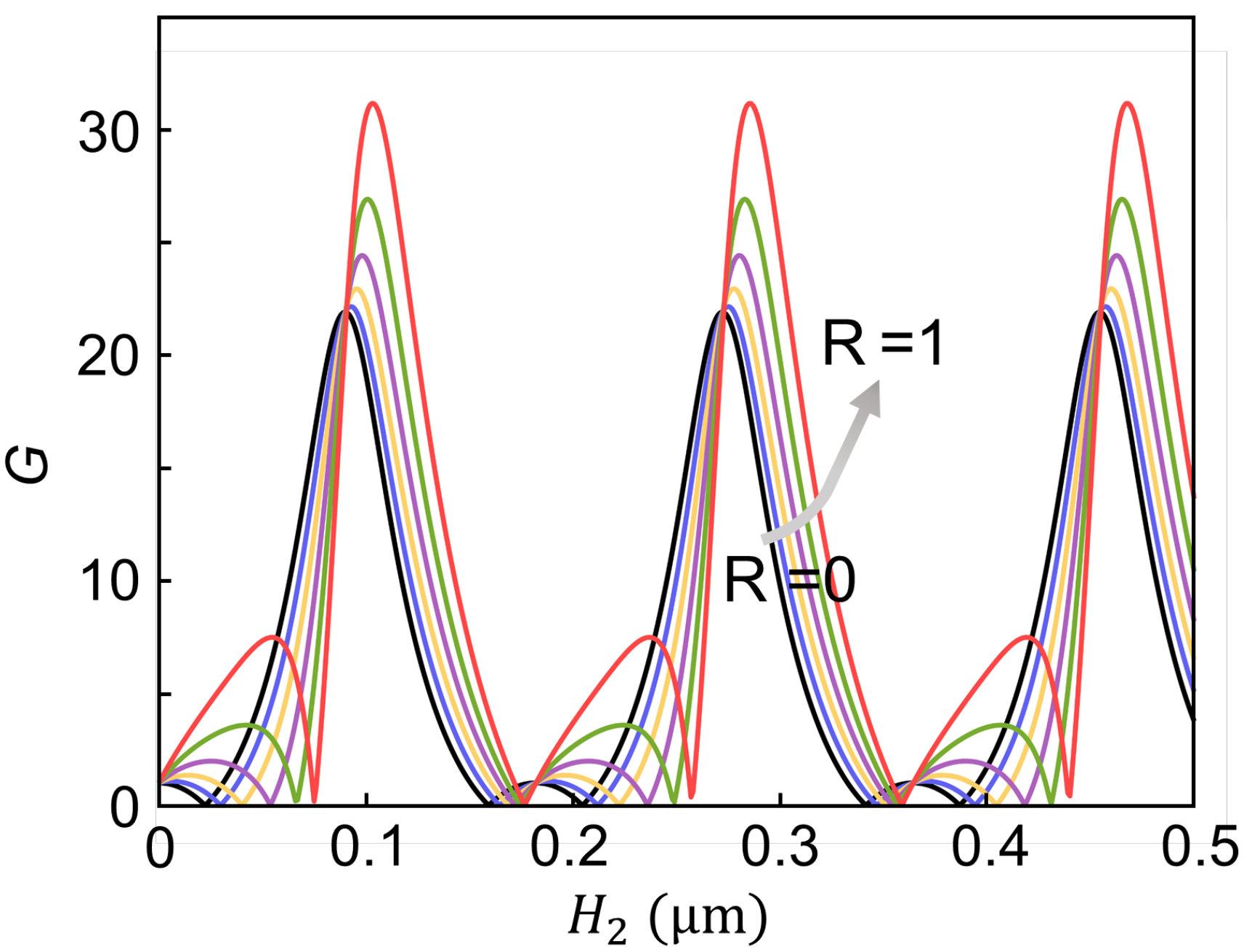

**Supplementary Fig. S7. The asymmetrical resonance in the phase cavity.** As the imaginary part becomes larger relative to the real part in the refractive index $\tilde{n}_1$ ($R$ from 0 to 1), the relationship between $G$ and cavity length $H_2$ becomes asymmetrical. Additionally, as the absorption of the sample layer increases (R increases), the upper limit of the phase gain increases.

## Supplementary Note 3. Comparison of wavelength-scanning experimental results using 50× and 20× objectives

We performed wavelength-scanning measurements on an MLG sample with a fixed $SiO_2$ thickness using 50x and 20x objectives separately to evaluate the effects of objectives on phase gain. The illumination wavelength was continuously tuned from 515 nm to 695 nm in 5 nm increments, enabling characterization of the phase-wavelength dependence, as summarized in **Fig. S8a**. To provide a more intuitive visualization of the changes in phase response under different illumination wavelengths, representative quantitative phase maps acquired at four selected wavelengths are displayed in **Fig. S8b**. These maps clearly illustrate the wavelength-dependent modulation of the measured phase contrast of the MLG sample.

Importantly, the phase-wavelength response curve obtained with the 50× objective exhibits the same trend as that measured using the 20× objective. This agreement indicates that, for structures with lateral dimensions exceeding the system's diffraction limit, the phase-wavelength relationship is governed by the intrinsic optical properties of the phase cavity rather than by the objective's magnification or numerical aperture. The result further verifies the reproducibility of the measurements and highlights the robustness of the wavelength-scanning phase measurements across different imaging configurations.

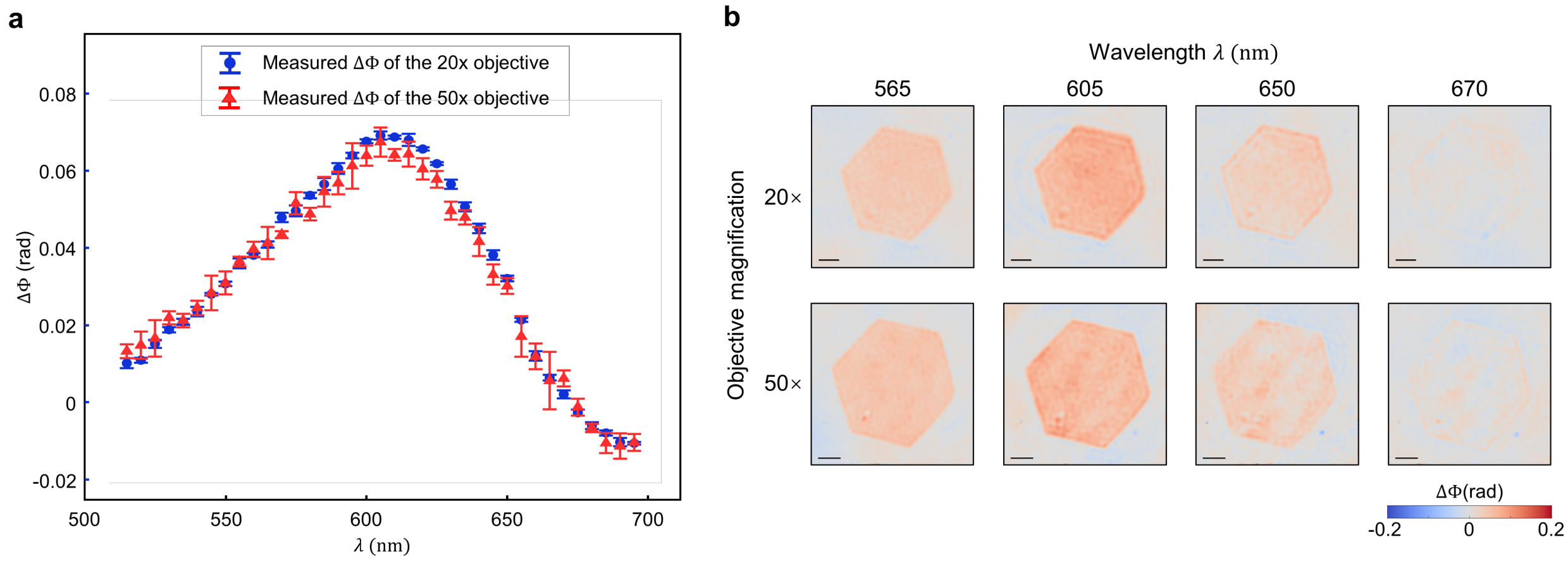

**Supplementary Fig. S8. Experimental verification of ΔΦ vs. λ using 20× and 50× objective lenses. a,** Measured ΔΦ vs. λ using 20× (blue) and 50× (red) objective lenses, separately. The wavelength scanning range is from 515 nm to 695 nm with an interval of 5 nm. **b,** Typical phase maps of the MLG sample acquired at λ = 565, 605, 650, and 670 nm are shown for both objectives. Scale bar: 5 μm.

**Supplementary Note 4. Simulations of phase gain with different cavity materials**

In addition to the $SiO_2$/Si structure, other commonly used wafers, including SiC/Si and $Si_3N_4$/Si can also function as single-layer phase cavities for phase amplification. We first simulated *G* as a function of wavelength $\lambda$ and SiC thickness for SiC/Si structure, as shown in **Fig. S9a.** At the widely used laser wavelength of $\lambda$ = 532 nm, the simulation shows that *G* reaches a maximum value of 12 on MLG under the resonance condition. Then, we simulated *G* as a function of wavelength $\lambda$ and $Si_3N_4$ thickness, as demonstrated in **Fig. S9b.** At $\lambda$ = 532 nm, the simulation shows that *G* reaches a maximum value of 930 on MLG under the resonance condition. From the simulation of $Si_3N_4$/Si structure in **Fig. S9b**, a high-precision tunable laser and sub-nanometer level thin film fabrication precision are required to meet the resonance condition for offering high phase amplification.

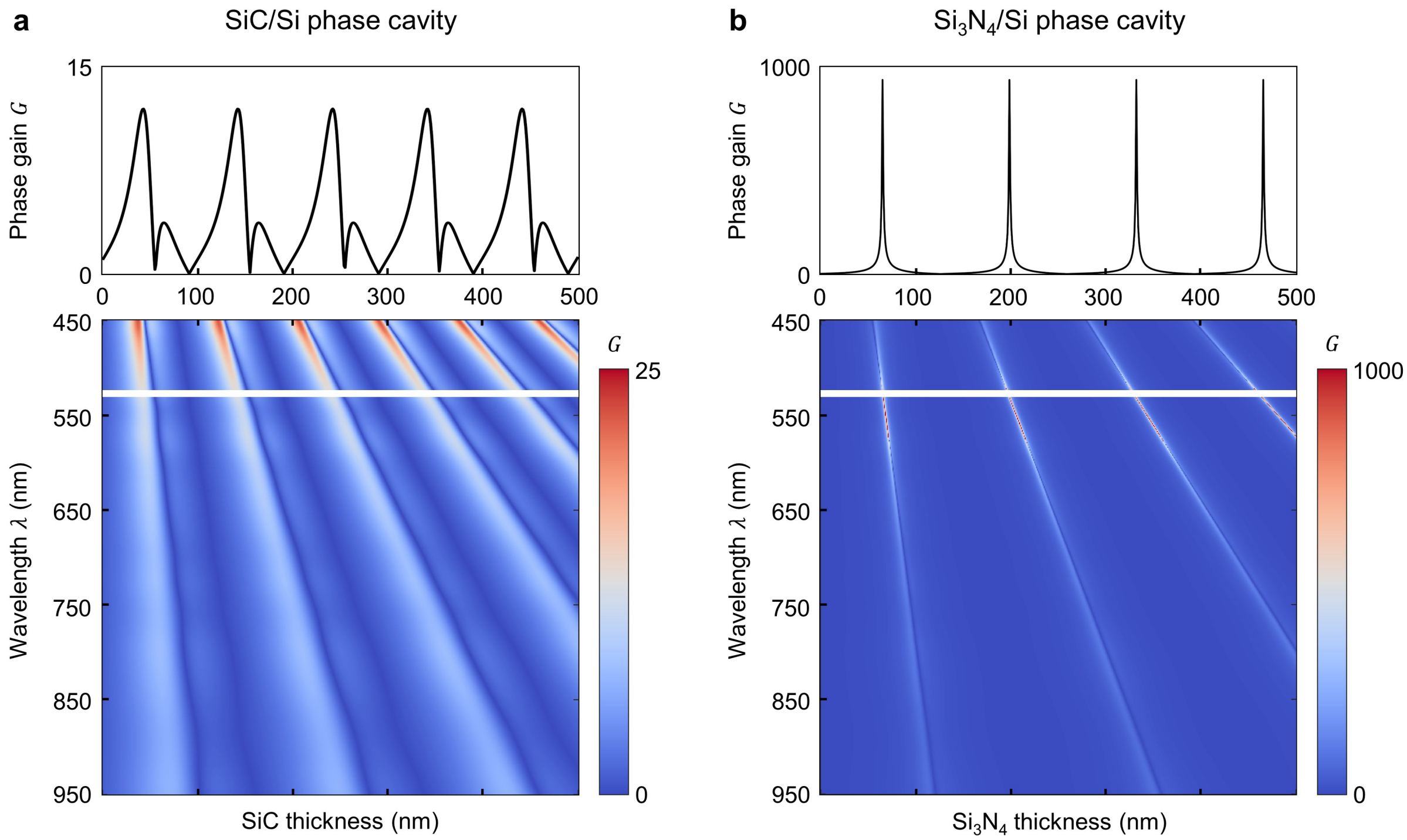

**Supplementary Fig. S9. Phase gain simulations of SiC/Si and $Si_3N_4$/Si single-layer phase cavities. a,** Phase gain simulation as a function of wavelength $\lambda$ and SiC thickness, where the periodic relation between $G$ and SiC thickness at 532 nm wavelength is plotted at the top. **b,** Phase gain simulation as a function of wavelength $\lambda$ and $Si_3N_4$ thickness, where the periodic relation between $G$ and $Si_3N_4$ thickness at 532 nm wavelength is plotted at the top.

**Supplementary Note 5. Phase gain demonstration of the multilayer phase cavity**

The phase gain mechanism can be generalized to different materials and further boosted by optimizing the cavity designs (e.g., multi-layer stacked cavity). When using the $SiO_2/Si_3N_4/Si$ design, as shown in **Fig. S10a**, we simulated the relationship between phase gain and the $SiO_2$ thickness $H_2$ and $Si_3N_4$ thickness $H_3$, when $\lambda$ = 532 nm. From the simulation result shown in **Fig. S10b**, it can be obtained that the phase gain of MLG mapping can reach 1306 when $H_2 = 178.7$ nm and $H_3 = 66.6$ nm. However, the resonance condition is more sensitive to the variations in layer thickness and laser wavelength than the single-layer phase cavity design. For example, a 0.1nm deviation from the thickness of $Si_3N_4$ to 66.6 nm may induce a gain reduction of 520. Moreover, we simulated the relationship between the amplitude of reflected wave $A$ and the $SiO_2$ thickness $H_2$ and $Si_3N_4$ thickness $H_3$, as shown in **Fig. S10c**. We found that the higher the phase magnification, the lower the amplitude of light wave $A$, resulting in the required laser power for imaging becoming tens or even hundreds of times higher. To experimentally validate the extended phase gain, we fabricated a $SiO_2/Si_3N_4/Si$ structure with $H_2 = 164.6$ nm and $H_3 = 69.3$ nm for mapping MLG, which corresponds to $G$ = 124 in the simulation. In experiments, we achieved 101.7-fold phase amplification on MLG, which is slightly lower than the phase gain. This difference could be attributed to the local cavity thickness variance from the fabrication inaccuracy, resulting in a mismatch from film thickness characterization. The phase imaging and thickness reconstruction results are shown in **Fig. S11**.

A detailed wavelength scanning results corresponding to the MLG sample shown in **Fig. 2h/3d** and the BLG sample shown in **Fig. 5c** are presented in **Fig. S12**.

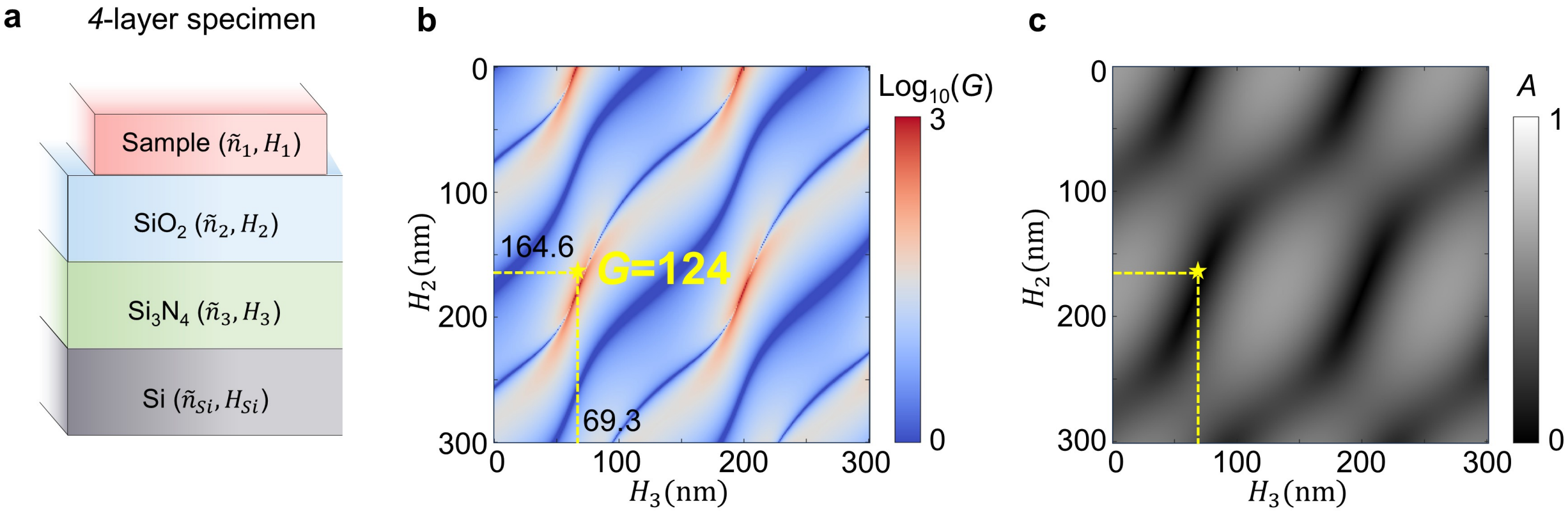

**Supplementary Fig. S10. a,** The schematic of a *4*-layer specimen consisting of the sample layer, $SiO_2$ layer, $Si_3N_4$ layer, and the reflective substrate Si layer. **b,** The simulation result of the relationship between phase gain $G$ and $SiO_2$ thickness $H_2$ and $Si_3N_4$ thickness $H_3$. **c**, The simulation result of the relationship between the amplitude of reflected wave $A$ and $SiO_2$ thickness $H_2$ and $Si_3N_4$ thickness $H_3$.

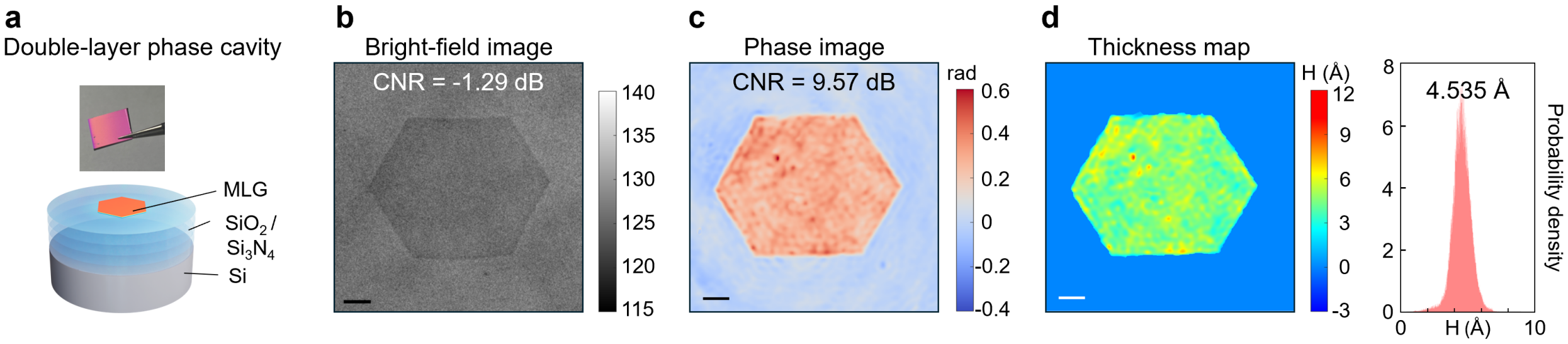

**Supplementary Fig. S11. Imaging results for MLG samples under a double-layer phase cavity. a,** The design and physical pictures of the single-layer phase cavity. **b,** The bright field image of MLG. **c,** The phase image of MLG. **d,** the thickness map and the corresponding histogram of MLG sample.

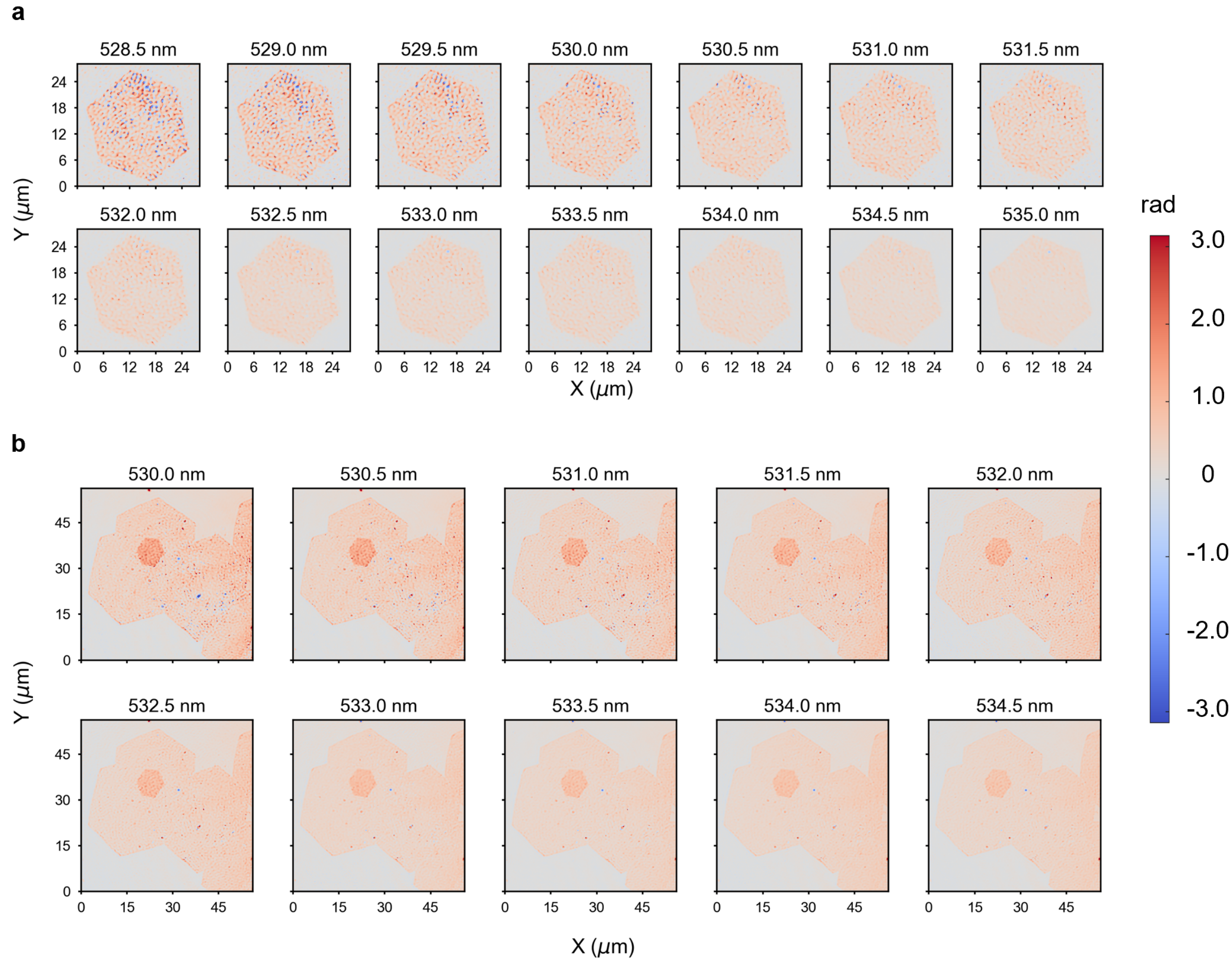

**Supplementary Fig. S12. The wavelength scanning result of MLG (a) and BLG (b) on the double-layer cavity.** Due to the strong absorption in the double-layer cavity and the limited power of the tunable laser, each image averages 100 frames with  an exposure of 2.5 second each.

**Supplementary Note 6. The thickness characterizations of thin films in phase cavities**

$SiO_2$/Si single-layer phase cavity characterization: The $SiO_2$ film thicknesses on silicon substrates were measured using a J.A. Woollam RC2 XI spectroscopic ellipsometer. The measured ellipsometric parameters (amplitude ratio $\Psi$ and phase difference $\Delta$) and their fitted curves are shown in **Fig. S13a-S13e**, in sequence. Five $SiO_2$/Si samples exhibited measured values of 260.3 nm, 278.5 nm, 285.8 nm, 308.3 nm, and 320.3 nm, respectively. Spectroscopic ellipsometry measurements were carried out over the wavelength range of 210–2500 nm, with spectral resolutions of 1 nm (210–1000 nm) and 6 nm (1000–2500 nm), at incident angles of 65° and 75°.

$SiO_2$/$Si_3N_4$/Si double-layer phase cavity: $Si_3N_4$ and $SiO_2$ layers were sequentially deposited on a bare silicon wafer using an Oxford Instruments Plasma Pro 800 Stratum plasma-enhanced chemical vapor deposition (PECVD) system. The $SiN_x$ layer was deposited for 231 seconds under gas flows of 500 sccm $N_2$, 11 sccm $NH_3$, and 300 sccm $SiH_4$ (5% in $N_2$) at 300°C, while the $SiO_2$ layer used 400 sccm $N_2$, 1000 sccm $N_2O$, and 400 sccm $SiH_4$ (5% diluted in $N_2$) at 300°C for 189s. Ellipsometric characterization of the resulting $SiO_2$/$Si_3N_4$ stack yielded thicknesses of $H_2$ = 164.6 nm ($SiO_2$) and $H_3$ = 69.3 nm ($Si_3N_4$), with the measured and fitted ellipsometry data in **Fig. S13f**.

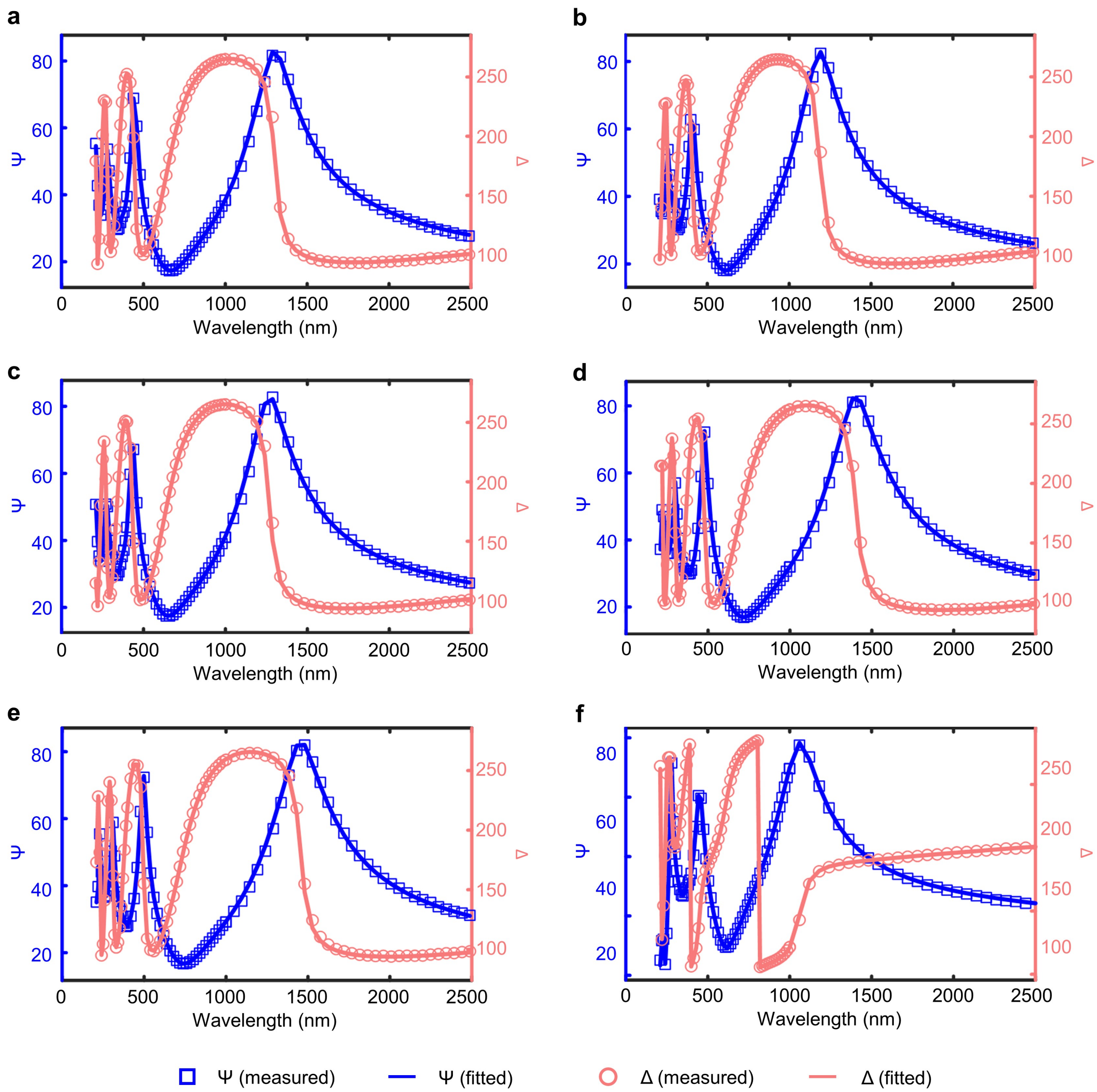

**Supplementary Fig. S13. Ellipsometry characterization of $SiO_2$ films and PECVD-deposited $SiO_2/Si_3N_4$ film stack. a-e**, Measured (symbols) and fitted (lines) ellipsometric parameters (Ψ, Δ) vs. wavelength at 65° incidence angle for five $SiO_2$ film samples with extracted oxide thicknesses of 260.3 nm, 278.5 nm, 285.8 nm, 308.3 nm, and 320.3 nm, respectively. **f**, Measured ellipsometric parameters (Ψ, Δ) data and fitted curves for the PECVD-deposited $SiO_2/Si_3N_4$ stack, with extracted thickness $H_2$ = 164.6 nm.

**Supplementary Note 7. Measurement accuracy derivation**

Based on the implicit relationship between the phase gain $G$ and the sample thickness $H_1$, we can define the measurement accuracy $\sigma_H$ using the system phase noise $\sigma_\phi$,

$$\sigma_H = \left| \left( \frac{\partial(\Delta\Phi)}{\partial H_1} \Big|_{H_1=0} \right)^{-1} \cdot \sigma_\phi \right|, \tag{S28}$$

where phase noise is not changed after phase amplification because the selective signal amplification provided by the phase cavity does not amplify this noise.

For a single-layer phase cavity, when $H_1$ is small (i.e., $|\varphi_1| = |2\pi n_1 H_1 / \lambda| \ll 1$, the reflected complex field $E_r$ is approximated into Eq. (S23) using the Taylor series expansion, $\Delta\Phi$ can be written as Eq. (S24), and $\Delta\varphi$ can be reduced to Eq. (S25). Combining with Eqs. (S22), (S24), and (S25), we can rewrite $\sigma_H$ as:

$$\sigma_H = \left( 2\pi \left| \mathrm{Re}\left\{ \frac{2\left(\tilde{n}_1^2 - n_0^2\right)}{\tilde{n}_{sub}^2 - n_0^2} \right\} \right| \right)^{-1} \times \frac{\sigma_\phi \lambda}{G} = \alpha \frac{\sigma_\phi \lambda}{G}. \tag{S29}$$

where $\alpha$ is a constant varying with the sample, and $\alpha = 0.307$ for graphene. It should be noted that Eq. (S29) is also applicable to a multilayer phase cavity case. As governed by Eq. (S29), the measurement accuracy of Φ-Amp is improved by amplifying the phase signal and reducing the phase noise.

The phase noise of conventional QPM is around 10 mrad under single-shot measurement condition, corresponding to a minimum detectable OPD in free space $\Delta OPD = \frac{\sigma_\phi \lambda}{2\pi} = 800$ pm. The minimum reported spatial noise with single-shot measurement is $\sigma_{\phi}$= $5.5\times10^{-4}$ rad, corresponding to $\Delta OPD$ = 47 pm[7]. However, the free-space OPD metric does not reflect practical measurement accuracy, as it neglects the influence of the sample's refractive index ( $\tilde{n}_1$ ). To translate the phase noise to measurement accuracy of graphene in a transmission-mode imaging system, we first retrieve the phase of the transmitted field in Eq. (S1). Then, we estimated the accuracy of 210 pm by using Eq. (S28), when the spatial phase noise $\sigma_{\phi}$= $5.5\times10^{-4}$ rad. The spatial noise was further reduced to $\sim 2.5\times10^{-4}$ rad through temporal averaging as reported by us and another group, respectively[7, 8], and it translates to an accuracy of ~100 pm. In this work, we break this accuracy limit by amplifying the phase signal (i.e., increasing *G*). As formalized in Eq. (S29), Φ-Amp can theoretically achieve femtometer-level accuracy, surpassing prior records by a factor of $10^3$, when applying a phase cavity with *G* of $10^3$.

Based on the relationship between the amplified phase, wavelength, cavity length, and the required measurand, we can assess the potential deviations in measurement accuracy caused by laser wavelength drift and fabrication deviations in the thin film thickness. By applying the error propagation formula under the first-order Taylor expansion, the deviation of measurement accuracy with respect to the variable $A$ ($A=\lambda, H_2, H_3$) is expressed as:

$$\delta H_1 = \left| \left( \frac{\partial \Phi}{\partial H_1} \Big|_{H_1=0} \right)^{-1} \cdot \frac{\partial \Phi}{\partial A} \delta A \right|. \tag{S30}$$

For a single-layer $SiO_2$/Si phase cavity, a laser wavelength drift of $\delta\lambda$ results in deviation $\delta H_1 \big|_{\lambda} = 2\times10^{-3}\delta\lambda$, and the fabrication deviation of $SiO_2$ thickness $\delta H_2$ corresponds to the deviation $\delta H_1 \big|_{H_2} = 2.2\times10^{-3}\delta H_2$. For a double-layer $SiO_2$/$Si_3N_4$/Si phase cavity, a laser wavelength drift of $\delta\lambda$ results in deviation $\delta H_1|_{\lambda} = 7.7\times10^{-3}\delta\lambda$, and the fabrication deviation of $SiO_2$ thickness $\delta H_2$ and $Si_3N_4$ thickness $\delta H_3$ corresponds to the deviation $\delta H_1|_{H_2} = 1.2\times10^{-2}\delta H_2$ and $\delta H_1|_{H_3} = 2.3\times10^{-2}\delta H_3$, respectively. For our current manufacturing process of phase cavities with nanoscale precision, the accuracy deviation for a single-layer cavity is 2.2 pm, but for a double-layer cavity, the accuracy deviation is as high as 23 pm. For a solid-state laser, the wavelength drifts with temperature by around 0.02 nm /℃. When the temperature is changed by 1 ℃, the accuracy deviation for a single-layer cavity is 0.04 pm, while for a double-layer cavity, the accuracy deviation is 0.154 pm. A narrowband laser illumination with a bandwidth of less than 5 nm can be adopted to further reduce coherent noise without causing too much accuracy deviation. For a double-layer cavity, such a bandwidth can induce an accuracy deviation of 38.5 pm. To sum up, both the laser and current manufacturing precision of phase cavities are insufficient for achieving measurement accuracy higher than the picometer level. A high-quality laser with a bandwidth of less than 0.01 nm is required to achieve an accuracy of tens of femtometers.

**Supplementary Note 8. Phase noise analysis**

To characterize the phase noise, we recorded 501 sample-free interferograms within 1 second. The first frame was used for calibration, and phase maps were extracted from the subsequent 500 frames. To quantify spatial phase noise, we calculated the average standard deviation across all 500 phase maps to minimize the impact of occasional measurement anomalies. This analysis yielded a spatial phase noise of 1.8 mrad. We further computed the frequency spectrum by averaging the spectra of individual pixels along the time domain (**Fig. S14a**), revealing two distinct peaks: one at around 50 Hz and the other spanning the 90 Hz- 100 Hz range. The noise at around 50 Hz mainly arises from AC power interference, while the 90 Hz- 100 Hz band corresponds to environmental disturbances (e.g., the mechanical vibrations).

After performing the temporal averaging with 100-frame summation, the spatial phase noise can be reduced to 0.7 mrad, suppressing photon shot noise and environmental noise. The theoretical photon shot noise of our system is 0.35 mrad, which can be theoretically reduced tenfold to 0.035 mrad after such temporal averaging. Therefore, we can estimate the noise contributions as shown in **Fig. S14b**, where the environmental noise contributes around 0.78 mrad and the speckle noise contributes around 0.67 mrad. It should be noted that the speckle noise is temporally correlated and thus it persists despite temporal averaging.

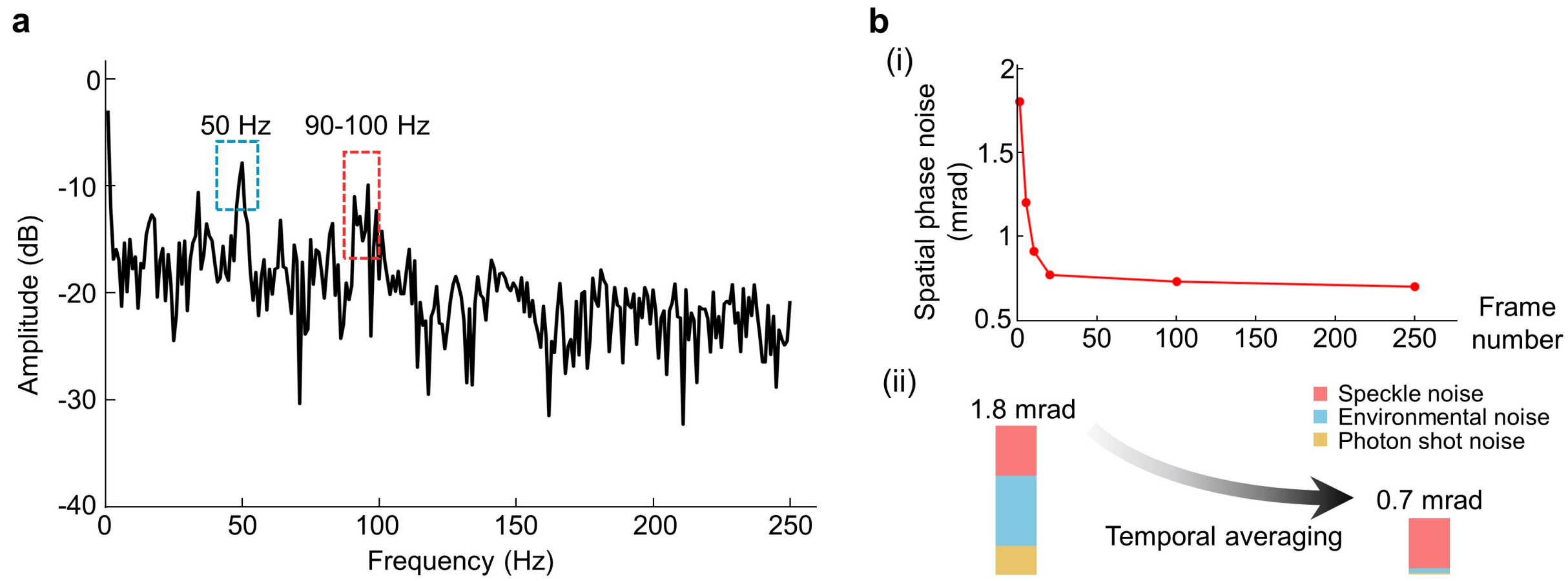

**Supplementary Fig. S14. a,** The averaged frequency spectrum (in logarithmic scale) of the measured time-series phase maps. **b,** (i) The relation between spatial phase noise with the frame number for summing. (ii) The noise contributions before and after temporal averaging.

**Supplementary Note 9. Density Functional Theory (DFT) for calculating the angle-dependent interlayer distance and refractive index of tBLG**

DFT is a quantum mechanical approach widely used to calculate the electronic structure and atomic arrangements of materials. DFT provides a framework for calculating the angle-dependent interlayer spacing of bilayer graphene by solving the Kohn-Sham equations:

$$\left[-\frac{\hbar^2}{2m}\nabla^2+V_{ext}\left(\mathbf{r}\right)+V_H\left(\mathbf{r}\right)+V_{XC}\left(\mathbf{r}\right)\right]\psi_i\left(\mathbf{r}\right)= {}_{i}\psi_i\left(\mathbf{r}\right), \tag{S31}$$

where $V_{ext}\left(\mathbf{r}\right)$ is the external potential resulting from nuclei, $V_H\left(\mathbf{r}\right)$ is the Hartree potential representing electron-electron Coulomb interactions, $V_{XC}\left(\mathbf{r}\right)$ is the exchange-correlation potential accounting for quantum effects, and $\psi_i\left(\mathbf{r}\right)$ and ${}_{i}$ are the Kohn-Sham orbitals and their corresponding energy eigenvalues, respectively.

For determining the interlayer spacing in bilayer graphene, the interactions between the two graphene layers, including van der Waals forces, can be accounted for through exchange-correlation functionals like the vdW functionals or dispersion corrections (e.g., DFT-D3). The interlayer spacing is obtained by minimizing the total energy $E_{tot}$:

$$E_{tot}\left[n\left(\mathbf{r}\right)\right]=T_s\left[n\right]+\int V_{ext}\left(\mathbf{r}\right)n\left(\mathbf{r}\right)d\mathbf{r}+E_H\left[n\right]+E_{XC}\left[n\right], \tag{S32}$$

where $n\left(\mathbf{r}\right)$ is the electron density. By systematically varying the distance between layers, the interlayer spacing is determined at the energy minimum, reflecting the balance of attractive and repulsive forces between the layers.

We used the open-source software Quantum ESPRESSO[9] for the *ab* initio calculation. To obtain reliable results, we performed convergence tests on pseudopotential-Van der Waals correction sets for the primitive cells of bulk graphite and AB-BLG. The predicted equilibrium interlayer distance values were compared with the well-accepted value for bulk graphite of 3.35 Å[10], as shown in **Supplementary Fig. S15.** Based on these tests, we adopted the Optimized Norm-Conserving Vanderbilt Pseudopotential (ONCVP) with Generalized Gradient Approximation Perdew-Burke-Ernzerhof (GGA-PBE)[11, 12] and DFT-D3 Van der Waals correction[13] for subsequent calculations. Under cell optimization convergence thresholds of $1\times10^{-8}$ Ry on total energy and $1\times10^{-6}$ Ry/Bohr on total force, we obtained an optimized in-plane lattice parameter of 2.4641 Å, consistent with values commonly referenced in tBLG research [14]. Following the modeling scheme for simulating tBLG with arbitrary twisting angles from prior studies[14], the bottom layer of graphene was treated as infinite to reduce the interactions between unit cells under periodic boundary conditions, while the upper layer was rotated with respect to its center to generate 4 sets of configurations with twisting angles ranging from 0° (AB-BLG) to 30° (30°-tBLG). The configurations of AB-BLG and 30°-tBLG are shown in **Fig. S16a**. A total of 338 atoms were included in a computational supercell. A vacuum layer with a thickness > 16 Å was introduced to further prevent interactions across unit cells.

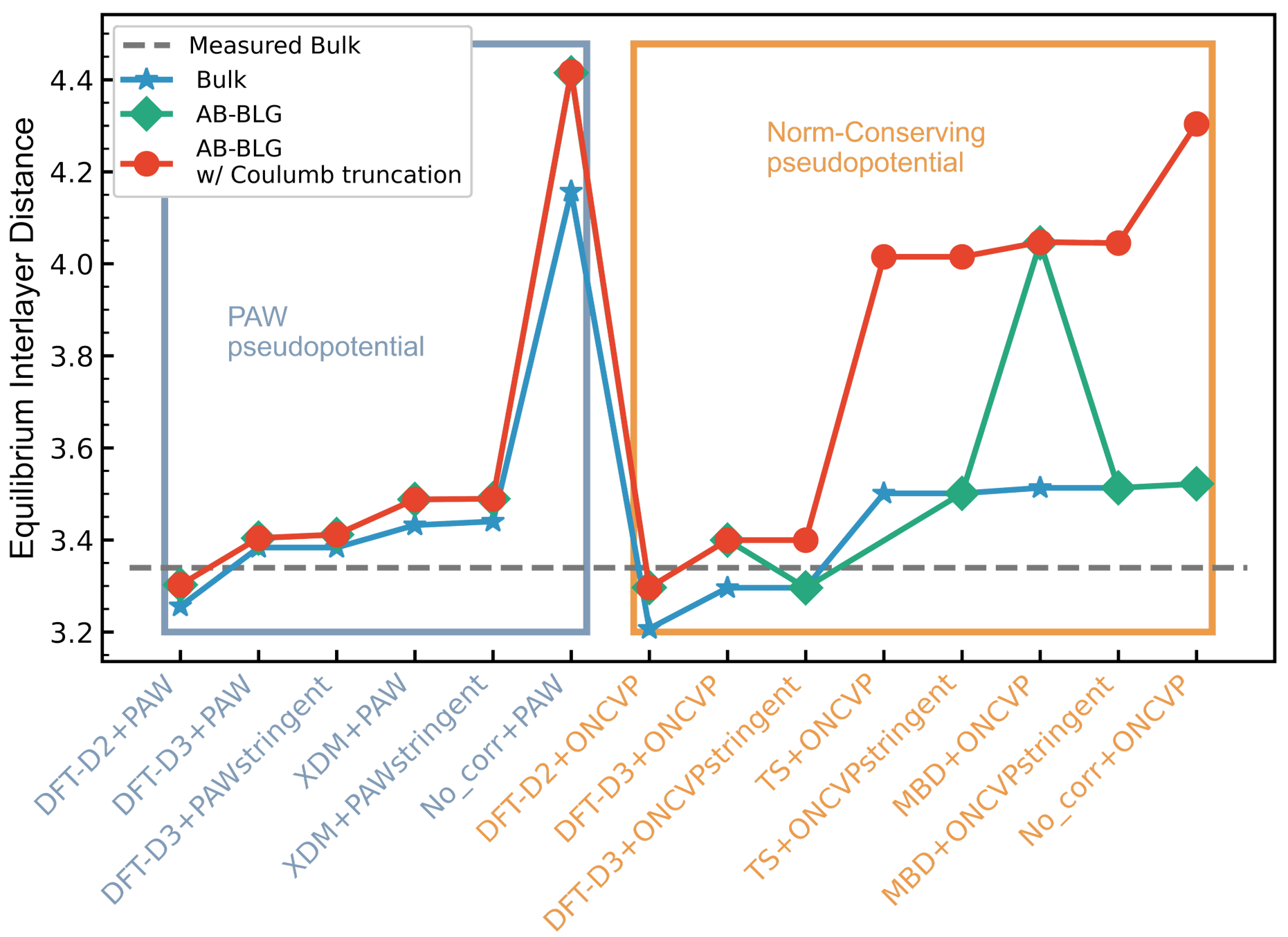

**Supplementary Fig. S15. Convergence tests of pseudopotential and van der Waals correction.** Sets of pseudopotentials and Van der Waals corrections were tested, taking the equilibrium interlayer distance of bulk graphite and AB-BLG as criteria. Pseudopotentials, including Projected Augmented Wave (PAW) and Optimized Norm-Conserving Vanderbilt Pseudopotential (ONCVP), were provided by PSEUDODOJO library. Van der Waals corrections used provided in pw.x package in Quantum ESPRESSO, including DFT-D2, DFT-D3, TS, MBD, XDM.

The equilibrium interlayer distances of 5 sets of configurations were determined by sweeping the distance between the two layers from 3.2 Å to 3.5 Å, followed by self-consistent field (SCF) calculations, as shown in **Fig. S16a**. All calculations were performed before assessing the convergence of **k**-point sampling, energy cutoff for wavefunctions, and energy cutoff for charge density and potential. Eventually, a 2×2×1 **k**-point mesh was employed with a wavefunction cutoff of 70 Ry and a charge density cutoff of 700 Ry. The predicted interlayer spacings of both AB-BLG and 30°-tBLG are shown in **Fig. S16b** and **Fig. S16c**. The discrepancy between DFT calculations and Φ-Amp measurements can be attributed to deviations in the interlayer binding energy, despite the improved long-range interaction modeling and a more accurate C6 coefficient provided by the DFT-D3 correction compared to the DFT-D2 correction [13]. Nevertheless, DFT-D3 remains the most reliable approach for predicting graphene configurations, outperforming conventional DFT, which requires the inclusion of long-range Van der Waals dispersion contributions[13, 15].

In advance of the structural parameters, the dielectric responses of MLG and BLG were revisited. However, thickness and dielectric function are strongly coupled and can be converted to each other, but cannot be obtained simultaneously in theory. The refractive index measurement of MLG was conducted by ellipsometry with the constant interlayer distance in graphite, 3.4 Å [16]. To provide a reliable description of the refraction characteristics of the tBLG systems, macroscopic dielectric response at a low energy range covering the laser wavelength used in experiments was predicted based on DFT and many-body perturbation theory. In the simplest case, random phase approximation (RPA) was implemented into the linear response theory, with the open-source code YAMBO [17]. Considering the computational difficulty of optical response in periodic low-dimensional systems, Coulomb potential cutting technique was adopted at the non-periodic direction to introduce the local field effect. In our cases, the non-periodic direction is parallel to the out-of-plane axis. Besides, to obtain a more general 2D dielectric response, a 2D model of infinite atomic layers with a 20 Å vacuum layer at the out-of-plane axis was used. For MLG and AB-BLG, the atomic models consisted of 2 and 4 atoms, respectively, with a 40 Ry wavefunction cutoff, 60×60×1 and 90×90×1 **k**-point mesh, respectively. For 30°-tBLG, the following modeling scheme was adopted to define one of the Moiré pattern lattice vectors:

$$\boldsymbol{L}_1 = m\boldsymbol{a}_1^{(1)} + n\boldsymbol{a}_2^{(1)} = n\boldsymbol{a}_1^{(2)} + m\boldsymbol{a}_2^{(2)}, \tag{S32}$$

by this, the twisting angle is defined by

$$\cos\theta = \frac{1}{2}\frac{m^2+n^2+4mn}{m^2+n^2+mn}, \tag{S34}$$

where $m$ and $n$ are integers, defining the ratio of the lattice vectors of two layers forming the Moiré pattern lattice vectors. However, the corresponding exact period of the Moiré pattern in 30°-tBLG is too large for computation. By choosing $m:n=1:3$, a model with a twisting angle of 32.2042° was obtained with 52 atoms, see the atomic model in **Fig. S17a**. Given the limited computational resources, this approximated model was selected to mimic the case of 30°-tBLG, with a 15×15×1 **k**-point mesh and 60 Ry wavefunction cutoff. Compared to interlayer distance calculation, these dielectric response simulations need a denser **k**-point mesh to capture as much as possible of electronic states. Under all of the above constraints, the DFT-predicted dielectric response at the in-plane direction of MLG and tBLG were shown in **Fig. S17b** in the form of 2D polarizability defined as

$$\alpha_{2D}(\omega)=\lim_{q\to 0}\frac{L}{4\pi \boldsymbol{q}^2}\chi_{00}(\boldsymbol{q},\omega), \tag{S35}$$

in which $L$ refers to the cell length in the non-periodic direction, providing the length gauge to the 2D polarizability. To obtain the macroscopic dimensionless dielectric response, the length gauge was eliminated by dividing the 2D polarizability by the actual thickness of each case and then obtaining the macroscopic electric susceptibility $\chi_{00}(\omega)$. Eventually refractive index is computed by $n(\omega)=\sqrt{\varepsilon(\omega)}=\sqrt{\chi_{00}(\omega)+1}$. Noteworthy, the current calculation contained only the electron-electron interaction of dielectric response. For further accuracy, the quasiparticle correction by GW approximation and electron-hole interaction by the Bethe-Salpeter Equation method should be involved in the optical calculation of MLG and tBLG. However, the above calculations are computationally expensive and out of our scope in this research. Results from several computational works [18] showed the less contribution of electron-hole interaction to the dielectric properties of BLG at 532 nm, corroborating that local RPA is enough for the present stage of optical properties prediction.

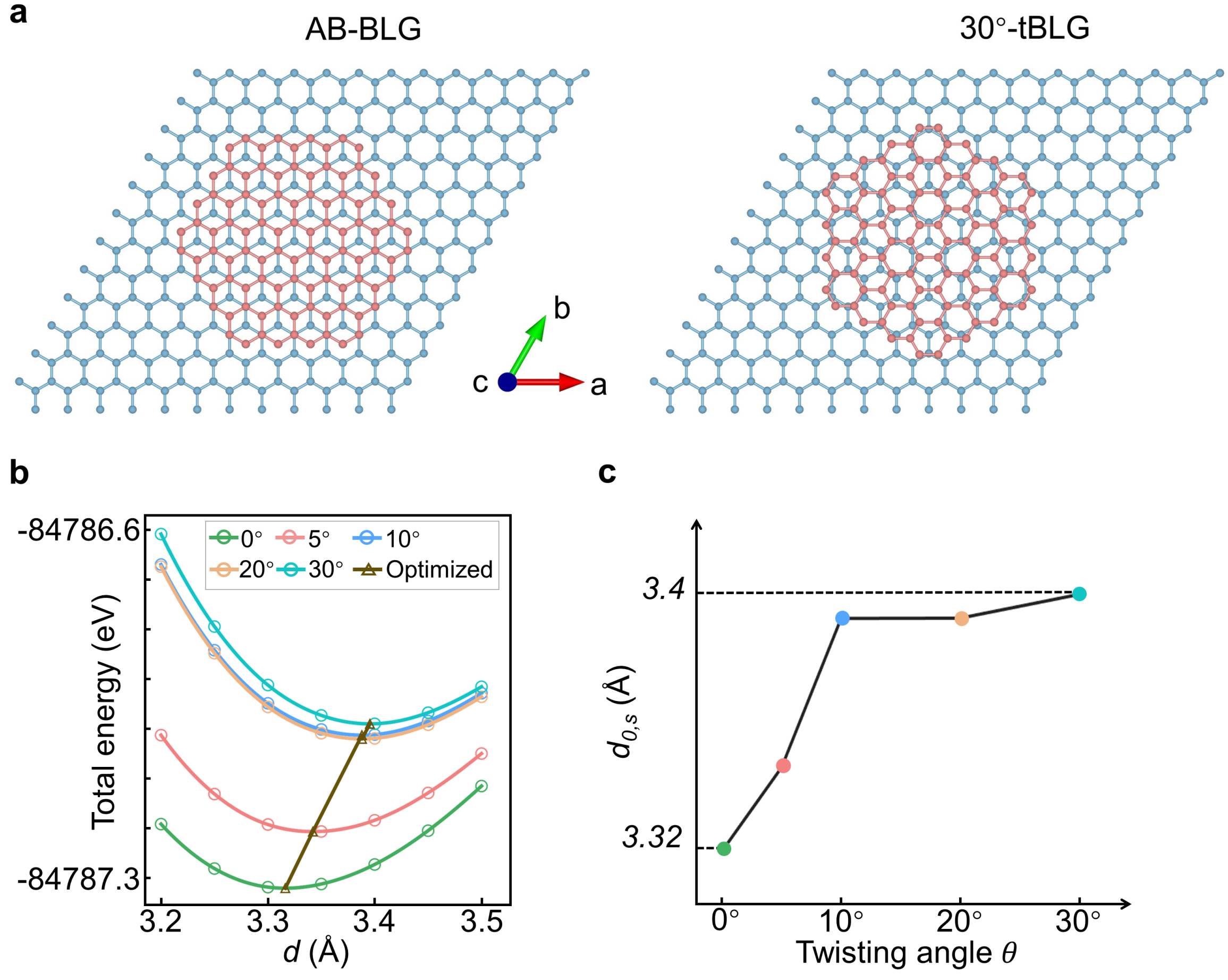

**Supplementary Fig. S16. DFT calculations for BLG interlayer distances with different twisting angles. a,** The used configurations of two BLG samples with twisting angle of 0° (AB-BLG) and 30° (30°-tBLG), respectively. **b,** DFT calculations result of total energy of BLG with different twisting angles, including $0°, 5°, 10°, 20°$ , and 30°. **c**, Optimal interlayer distance of tBLG changes as twisting angle changes.

The error estimation of DFT predicted refractive index was mainly focused on the explicit expression of the refractive index regarding the actual thickness. As mentioned in the main text, the coupled thickness and dielectric response hindered the progress of the thickness measurement. Here, a guess of the atomic layer thickness of BLG was needed. It led to an expression of refractive index based on DFT predicted 2D polarizability as

$$n(\omega)=\sqrt{\frac{4\pi\alpha_{2D}(\omega)}{H}+1}=\sqrt{\frac{2\pi\alpha_{2D}(\omega)}{d}+1}, \tag{S36}$$

For BLG, the total thickness $H$ is the double of atomic layer thickness $d$. Expanding Eq. (S36) and taking the first-order derivative at $d$ = 3.4 Å, the refractive index error caused by the error of atomic layer thickness was defined as

$$\Delta n(\omega)\Big|_{d=3.4} = \frac{\partial n(\omega)}{\partial d}\Bigg|_{d=3.4} \Delta d = -\frac{1}{2}\frac{\pi\alpha_{2D}(\omega)}{d^2\sqrt{\dfrac{2\pi\alpha_{2D}(\omega)}{d}+1}}\Bigg|_{d=3.4} \Delta d\,, \tag{S37}$$

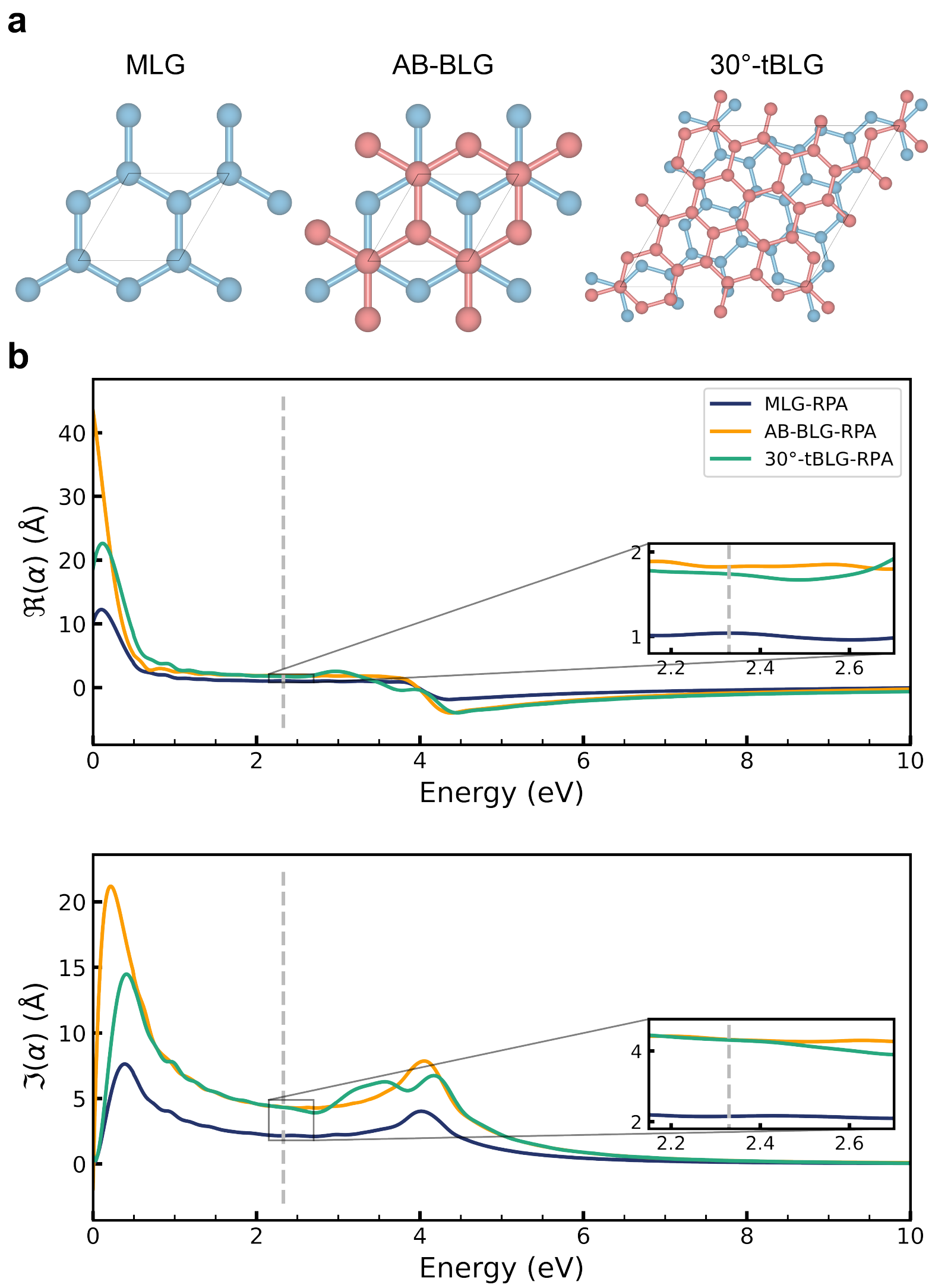

**Supplementary Fig. S17. DFT and many-body perturbation theory calculations of MLG and BLG dielectric responses. a**, The used configurations of MLG, AB-BLG, and 30°-tBLG, respectively. **b**, local field RPA calculations result of 2D polarizability of MLG, AB-BLG, and 30°-tBLG, in terms of real part and imaginary part.

In the experiment conducted in **Fig. 5b**, we further repeated the measurements using three independent samples for each stacking order and conducted the statistical analysis of the interlayer spacing difference. As shown in **Fig. S18a**, an independent-sample *t*-test was conducted to compare the differences between the two groups, and we obtained a $p$-value <0.02, confirming a statistically significant difference in interlayer spacing between the two configurations. The 95% confidence interval for the interlayer spacing difference is 0.77±0.50 Å.

Using DFT calculations, we also estimated the interlayer spacings of AB-BLG and 30°-tBLG of 3.32 Å and 3.40 Å, which are underestimated due to the restriction of the approximated correction. In the thickness reconstruction, we need accurate and reliable refractive index values of AB-BLG and 30°-tBLG and their MLG regions, so we calculated them using DFT and linear response theory with local field Random Phase Approximation (RPA)[17] (**Fig. S18b**), and the values are consistent with the literature[19]. The oscillations observed in **Fig. S18b** arise from the finite **k**-point sampling used in the calculations. To verify the stacking orders of tBLG, we obtained the Raman spectra of the same AB-BLG and 30°-tBLG samples used in Φ-Amp mapping, as shown in **Fig. S18c**. The full width at half maximum (FWHM) of the 2D peak in AB-BLG is significantly larger than MLG and 30°-tBLG, which are consistent with the previous report[20].

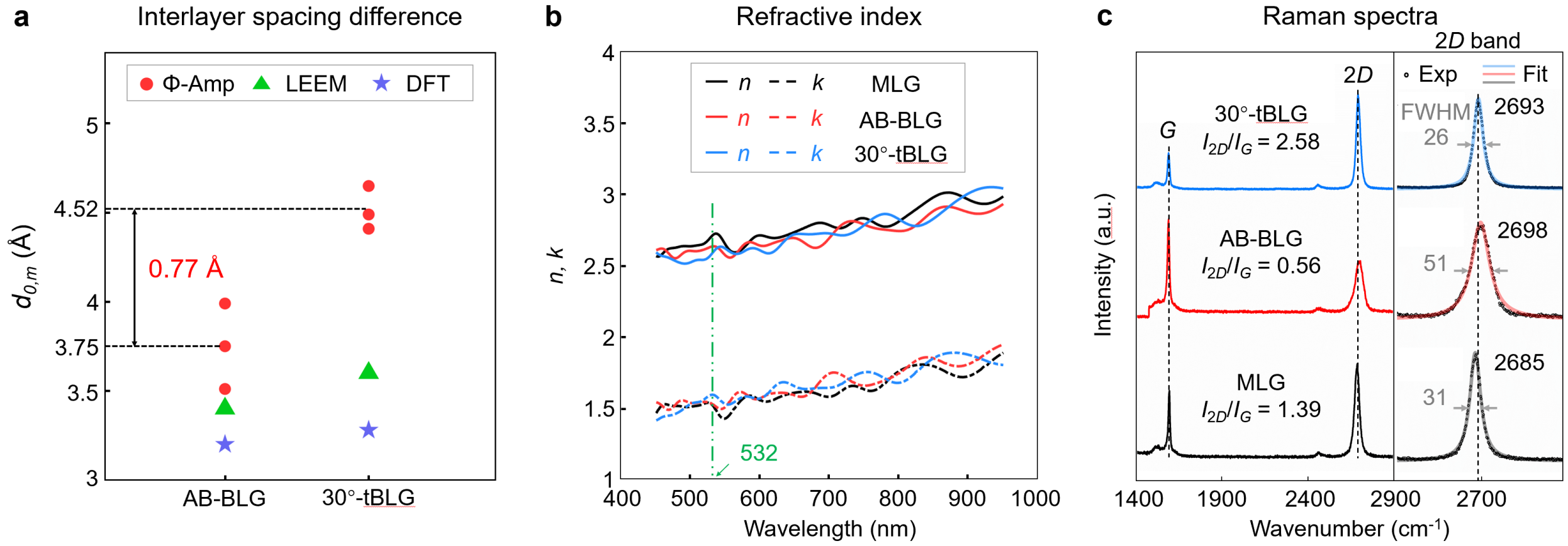

**Supplementary Fig. S18. Interlayer spacing analysis of AB-BLG and 30◦-tBLG. a,** Independent-samples *t*-test shows a 0.77 Å interlayer spacing difference between AB-BLG and 30°-tBLG ($p < 0.02$). The trend matches with LEEM measurements (triangles) and our DFT calculations (stars). **b,** Simulated RI spectral distributions of MLG, AB-BLG, and 30°-tBLG. **c,** Raman spectra, and Lorentz fitting of the 2D bands of MLG, AB-BLG, and 30°-tBLG, respectively.

**Supplementary Note 10. The influence of detection well capacity on thickness mapping accuracy**

In this study, we captured interferograms using a high full-well-capacity (HFWC) camera with an FWC of $2 \times 10^6$ electrons. For comparison, we also performed thickness measurements of MLG using a regular camera with a FWC of $1 \times 10^4$ electrons (MV-CE013-80UM, Hangzhou Hikrobot Co., LTD.). The reconstructed thickness maps of the same MLG flake, captured by the HFWC camera and the regular camera, are shown in **Fig. S19a** and **Fig. S19b,** respectively. The corresponding thickness histograms in **Fig. S19c** indicate comparable measured thickness for both cameras (4.18 Å vs. 4.44 Å). We also calculated CNR values of the HFWC camera image of 7.14 dB, while the regular camera image exhibited a lower CNR of 5.05 dB. This demonstrates that though both cameras can detect MLG thickness, the regular camera introduces higher detection noise than an HFWC camera, according to phase sensitivity theory.

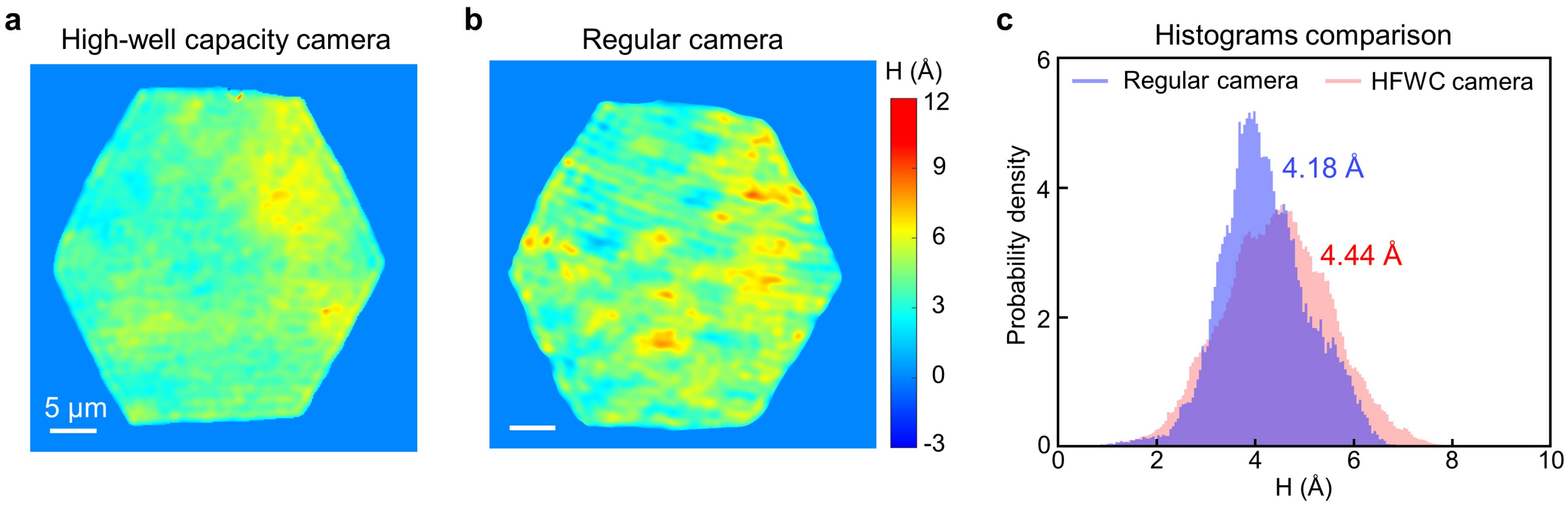

**Supplementary Fig. S19. The reconstructed thickness maps of MLG captured using different cameras. a,** High well-capacity camera. **b,** Regular camera. **c,** Thickness histograms comparison of MLG in **a** and **b**.